# Determining the Neutrino Mass Eigenstates and the effective Majorana Mass

Zoran B. Todorovic

Faculty of Electrical Engineering, Department of Physics, University of Belgrade, Belgrade, Serbia

Email address:tzoran221@gmail.com

**Abstract**: This paper aims at solving several open questions in current neutrino physics: the neutrino mass hierarchy, the Dirac CP violating phase, the absolute mass of neutrinos, the nature of neutrinos (Dirac or Majorana), the Majorana matrix and the absolute value of the effective Majorana neutrino mass.In the research presented in this paper, we have shown that the precise definition of the mass splittings between neutrino mass eigenstates, done in the latest analysis of experimental data, can be of crucial importance for defining the nature of neutrino mass hierarchy. The Standard Model has three generations of fundamental matter particles. Three generations of the charged lepton mass show a hierarchical structure: $m_\tau > m_\mu > m_e$. Owing to that, there is a belief and it is considered that neutrinos may follow such hierarchical structure. In our calculations, we have also included the latest data obtained, based on the processing of measurement results, which showed that even with such data, obtained results favor the normal neutrino mass hierarchy. As for the individual neutrino mass calculated in this paper, in today's neutrino physics it is only known that neutrino mass scale is bounded only from above, and both the Dirac and the Majorana character of neutrinos are compatible with all observations. Among some of the questions resolved in this paper, which are related to the properties of neutrinos, a positive answer was also given to the question of whether light neutrinos are self-conjugate particles or not.

1. Introduction

This paper presents models of neutrinos selected on the basis of their affiliation to one of the two possible hierarchies of neutrino masses. The selection was made on the basis of strict and exclusive affiliation to one of the possible hierarchies of neutrino masses.
Four examples were analysed, the first two of which belong exclusively to the inverse hierarchy of neutrino masses and the third and fourth examples belong exclusively to the normal hierarchy of neutrino masses. These models indicate the areas in which neutrino masses could be found. However, we have further analysed these models with the intention of determining the positions in these areas where explicit numerical values of neutrino masses should be found.

In neutrino physics, based on the assumption that neutrinos are particles of Majorana by nature a formula for possible value for Majorana neutrino mass is derived [1].

Numerous values for neutrino masses, Majorana phases and Dirac CP violation phase is present in this formula.

However, for all these physical quantities, areas in which they could be found are still shown, which has the consequence that the value for the Majorana mass itself is presented as possible values that could be found in the corresponding area.

This uncertainty in the explicit numerical values of these physical quantities was the reason for developing a procedure to find points in these areas related to their numerical values.

In cosmology, an upper limit has been set for the sum of neutrino masses, which reads: $\sum m_i < 0.12 eV$ .

Thus, one comes to the conclusion that, due to that limit, the normal hierarchy of neutrino masses would be favored in relation to the inverse neutrino mass. Therefore, all further research in this paper has been subjected to this criterion through the procedure of calculating and defining some of the properties of neutrinos, such as:

1. The Dirac CP violation phase $\delta_{CP}$ .
2. The absolute mass of a neutrino.
3. The Majorana phases.
4. Do neutrinos and antineutrinos behave differently? Is a neutrino its own antiparticle? Or in other words: are they Dirac or Majorana particles?
5. What is the absolute value of the effective Majorana neutrino mass?
6. The Majorana matrix.

## 2. The neutrino mass sum rule

In order to determine the mass sum rule, we start from the general form of the neutrino mass sum rule given in Ref. [1]:

$$A_1 m_1^p e^{i\tilde{\phi}_1} + A_2 m_2^p e^{i\tilde{\phi}_2} + A_3 m_3^p e^{i\tilde{\phi}_3} = 0 \tag{1}$$

The geometric interpretation of the relation (1) is shown in the complex plane in the shape of a triangle with sides described in the form of complex numbers:

$$m_1^p e^{i\tilde{\phi}_1}, m_2^p e^{i\tilde{\phi}_2}, m_3^p e^{i\tilde{\phi}_3} \tag{2}$$

The equation (1) is simplified by multiplying it with $e^{-i\tilde{\phi}_1}$ , while all the parameters are selected in the following manner:

$$\tilde{\phi}_1 = p\phi_1 + \chi_1, \tilde{\phi}_2 = p\phi_2 + \chi_2, \tilde{\phi}_3 = p\phi_3 + \chi_3 \tag{3}$$

Where the parameter *p* in relation to (3) could have the following values: $p = \pm 1/2, \pm 1/3, \pm 1/4, \pm 1$ The phases $\chi_i$ are not Majorana phases, $[0, 2\pi)$, which are fixed phases, and they are chosen according to the selected model of the mass sum rule [1], $\phi_i, i = 1,2,3$ are Majorana phases appearing in the process of neutrinoless double beta decay, and they are explained with the Feynman diagrams for the process of obtaining effective neutrino Majorana mass [1]. Based on the relations (3) phase differences are formed

$$\tilde{\phi}_2 - \tilde{\phi}_1 = \phi_2 - \phi_1 + \chi_2 - \chi_1 = p\alpha_{21} + \Delta\chi_{21}; \alpha_{21} = \phi_2 - \phi_1, \tilde{\phi}_3 - \tilde{\phi}_1 = \phi_3 - \phi_1 + \chi_3 - \chi_1 = p\alpha_{31} + \Delta\chi_{31}; \alpha_{31} = \phi_3 - \phi_1.$$ (4)

And they are included in the triangle equation (1) assuming the following form:

$$A_1 m_1 + A_2 \left(m_2 e^{i\alpha_{21}}\right)^p e^{i\Delta\chi_{21}} + A_3 \left(m_3 e^{i\alpha_{31}}\right)^p e^{i\Delta\chi_{31}} = 0$$ (5)

## 2.1 *The first example. Group* $A_5'$ *,Seesaw Type Weinberg, Matrix* $M_\nu$

Here we will use the data published in Ref. [3]:

$$\Delta m_{21}^2 = (7.53 \pm 0.18) \times 10^{-5} eV^2, \Delta m_{32}^2 = (2.453 \pm 0.033) \times 10^{-3} eV^2 (NO), \Delta m_{32}^2 = (-2,536 \pm 0.034) \times 10^{-3} eV^2 (IO),$$
$$\sin^2\theta_{12} = 0.307 \pm 0.013, \sin^2\theta_{23} = 0.539 \pm 0.0222 (IO), \sin^2\theta_{23} = 0.546 \pm 0.021 (NO), \sin^2\theta_{13} = (2.20 \pm 0.07) \times 10^{-2}.$$ (6)

In Ref. [1], several examples of neutrino mass rules have been analysed. For the purposes of our further research, a selection was made between those examples and only those that exclusively belong to only one hierarchy of neutrino masses were taken. Four examples were analysed, the first two of which belong exclusively to the inverse hierarchy of neutrino masses and the third and fourth examples belong exclusively to the normal hierarchy of neutrino masses. These neutrino models indicate the areas in which neutrino masses could be found. However, we have further analysed these models with the intention of determining the positions in these areas where explicit numerical values of neutrino masses should be found.

Among others, an example of the Weinberg operator for the light neutrinos is given, which is derived based on $A_5'$ symmetry group, and it is stated in the form:

$$\tilde{m}_1 - \tilde{m}_2 \frac{\sqrt{3}-1}{2} + \tilde{m}_3 \frac{\sqrt{3}+1}{2} = 0; \tilde{m}_j = m_j \exp(i\phi_j), j = 1,2,3.$$ (7)

The identification of formulas (5) and (6) shows that:

$$p = +1, \Delta\chi_{21} = \pi, \Delta\chi_{31} = 0, B_2 = \frac{\sqrt{3}-1}{2}, B_3 = \frac{\sqrt{3}+1}{2}.$$ (8)

and we especially emphasize that relation (7) belongs to the inverted mass hierarchy: $m_3 < m_1 < m_2$. With this note, and bearing in mind that neutrino masses are real quantities, the mass sum rule for the Weinberg operator takes the form of

$$m_1 - m_2 \exp(i\alpha_{21})\frac{\sqrt{3}-1}{2} + m_3 \exp(i\alpha_{31})\frac{\sqrt{3}+1}{2} = 0 \tag{9}$$

Equation (8) is complex, and can be separated into a real part and an imaginary part. The real part is described by the equation:

$$m_1 - m_2 \cos(\alpha_{21})\frac{\sqrt{3}-1}{2} + m_3 \cos(\alpha_{31})\frac{\sqrt{3}+1}{2} = 0 \tag{10}$$

And the imaginary part is the equation

$$-m_2 \sin(\alpha_{21})\frac{\sqrt{3}-1}{2} + m_3 \sin(\alpha_{31})\frac{\sqrt{3}+1}{2} = 0 \tag{11}$$

As it is obvious, equation (11) is satisfied for all solutions related to values:

$$\alpha_{21} = 0, \pm\pi, \pm 2\pi, \alpha_{31} = 0, \pm\pi, \pm 2\pi. \tag{12}$$

Here it is important to point out that $\alpha_{21}, \alpha_{31}$ are Majorana phases that have physical sense, and which appear in the process of neutrinoless double beta decay [2] [4] [5] [6] [7]. If we take equations into account, (10) and (11), we will reach a unique solution:

$$\alpha_{21} = 0, \alpha_{31} = \pi. \tag{13}$$

Taking into account solutions (13), equation (10) becomes

$$m_1 = m_2 \frac{\sqrt{3}-1}{2} + m_3 \frac{\sqrt{3}+1}{2} \tag{14}$$

We have one equation with three unknowns, $m_1, m_2, m_3$, and we will solve these three unknowns by performing the following procedure. In the first step, by squaring the left and right sides of this equation, we get the equation which reads:

$$-m_1^2 + m_2^2 - \frac{\sqrt{3}}{2}m_2^2 + m_3^2 + \frac{\sqrt{3}}{2}m_3^2 + m_2 m_3 = 0 \quad (15)$$

This equation gets its full physical sense because there are measured values [3]

$$\left(\Delta m_{21}^2 = m_2^2 - m_1^2, \Delta m_{13}^2 = m_1^2 - m_3^2, \Delta m_{23}^2 = m_2^2 - m_3^2\right)(IO), \left(\Delta m_{21}^2 = m_2^2 - m_1^2, \Delta m_{31}^2 = m_3^2 - m_1^2, \Delta m_{32}^2 = m_3^2 - m_2^2\right)(NO) \quad (16)$$

Taking into account (16), we can represent equation (15) by two equations. The first one is

$$m_3^2 + m_2 m_3 - C = 0, C = \frac{\sqrt{3}}{2}\Delta m_{23}^2 - \Delta m_{21}^2. \quad (17)$$

And the second one is

$$m_2^2 + m_2 m_3 - D = 0, D = \frac{\sqrt{3}}{2}\Delta m_{23}^2 + \Delta m_{13}^2. \quad (18)$$

The solution of equation (17) as a quadratic equation is

$$m_3 = -\frac{m_2}{2} \pm \sqrt{\left(\frac{m_2}{2}\right)^2 + C} \quad (19)$$

We insert this solution for $m_3$ into equation (18) thus obtaining an equation of unknown magnitude $m_2$:

$$\left[\frac{m_2^2}{2} - D\right] = \left[\mp m_2 \sqrt{(m_2/2)^2 + C}\right] \quad (20)$$

We find a solution for a numerical value, for $m_2$:

$$m_2 = \frac{D}{\sqrt{C+D}}, = \frac{\sqrt{3}/2\Delta m_{23}^2 + \Delta m_{13}^2}{\sqrt{\sqrt{3}\Delta m_{23}^2 + \Delta m_{13}^2 - \Delta m_{21}^2}} = \frac{\langle U \rangle}{\sqrt{\langle V \rangle}} + |\Delta m_2| = \frac{\sqrt{3}/2 \times (0.002536 \pm 0.000034) + (0.0024607 \pm 0.0000358)}{\sqrt{\sqrt{3} \times (.002536 \pm 0.000034) + (0.0024607 \pm 0.0000358) - (0.0000753 \pm 0.0000018)}} = \langle m_2 \rangle + |\Delta m_2|,$$

$$\langle m_2 \rangle = \frac{\langle U \rangle}{\sqrt{\langle V \rangle}} = \frac{\sqrt{3}/2 \times (0.002536) + 0.0024607}{\sqrt{\sqrt{3} \times 0.002536 + 0.0024607 - 0.0000753}} = 0.056565771797\, eV/c^2, |\Delta m_2| = \left|\frac{\partial \left(\frac{\langle U \rangle}{\sqrt{\langle V \rangle}}\right)}{\partial \langle U \rangle}\right| |\Delta U| + \left|\frac{\partial \left(\frac{\langle U \rangle}{\sqrt{\langle V \rangle}}\right)}{\partial \langle V \rangle}\right| |\Delta V| = \frac{1}{\sqrt{\langle V \rangle}} |\Delta U| + \frac{1}{2}\frac{\langle U \rangle}{\sqrt{\langle V \rangle}} \left|\frac{1}{\langle V \rangle}\right| |\Delta V| \qquad (21)$$

$$= \frac{1}{\sqrt{\langle V \rangle}} |\Delta U| + \frac{1}{2}\langle m_2 \rangle \left|\frac{1}{\langle V \rangle}\right| |\Delta V|, |\Delta U| = \sqrt{3}/2 \times 0.000034 + 0.0000358, |\Delta V| = \sqrt{3} \times 0.000034 + 0.0000358 + 0.0000018.$$

Applying the previous procedure we arrive to the solution for a numerical value for $m_3$:

$$m_3 = \frac{C}{\sqrt{C+D}}, \qquad (22)$$

Thus, we obtained solutions for neutrino mass eigenstates as a function of the measured parameters $\Delta m_{21}^2, \Delta m_{23}^2, \Delta m_{13}^2$.

*Numerical values of neutrino mass*

$$m_3 = \frac{C}{\sqrt{C+D}} = \frac{U}{\sqrt{V}} = \langle m_3 \rangle + |\Delta m_3| = \frac{\sqrt{3}/2 \times (0.002536 \pm 0.000034) - (0.0000753 \pm 0.0000018)}{\sqrt{\sqrt{3} \times (0.002536 \pm 0.000034) + (0.0024607 \pm 0.0000358) - (0.0000753 \pm 0.0000018)}} = (0.025762 \pm 0.000563) eV/c^2,$$

$$\langle m_3 \rangle = \frac{\langle U \rangle}{\sqrt{\langle V \rangle}} = \frac{\sqrt{3}/2 \times (0.002536) - 0.0000753}{\sqrt{\sqrt{3} \times 0.002536 + 0.0024607 - 0.0000753}} = 0.025762114413552 \, eV/c^2,$$

$$|\Delta m_3| = \left|\frac{\partial \left(\frac{\langle U \rangle}{\sqrt{\langle V \rangle}}\right)}{\partial \langle U \rangle}\right| |\Delta U| + \left|\frac{\partial \left(\frac{\langle U \rangle}{\sqrt{\langle V \rangle}}\right)}{\partial \langle V \rangle}\right| |\Delta V| = \frac{1}{\sqrt{\langle V \rangle}} |\Delta U| + \frac{1}{2} \frac{\langle U \rangle}{\sqrt{\langle V \rangle}} \frac{1}{\langle V \rangle} |\Delta V| = \frac{1}{\sqrt{\langle V \rangle}} |\Delta U| + \frac{1}{2} \langle m_3 \rangle \frac{1}{\langle V \rangle} |\Delta V|, \quad (23)$$

$$|\Delta U| = \sqrt{3}/2 \times (0.000034) + (0.0000018), \quad |\Delta V| = \sqrt{3} \times (0.000034) + (0.0000358) + (0.0000018),$$

$$m_2 = \frac{D}{\sqrt{C+D}} = (0.0565658 \pm 0.0011952) eV/c^2,$$

$$m_1 = \frac{\sqrt{3}+1}{2} \frac{C}{\sqrt{C+D}} + \frac{\sqrt{3}-1}{2} \frac{D}{\sqrt{C+D}} = (0.0558962 \pm 0.0012064) eV/c^2.$$

In cosmology, the possible limit value for the sum of neutrino masses is measured and it amounts to $\sum m_i < 0.12 eV$. In this case we find the sum of masses $\sum m_i \approx (0.138 \pm 0.003) eV$ which does not fit the upper limit in cosmology.

## 2.2 The second example. Group $S_4$, Seesaw type Dirac, Matrix $M_\nu$

In the Ref. [1], several examples of the neutrino mass rules have been analysed. Among others, the example of the Dirac neutrinos from flavour symmetry $S_4$ group for the light neutrinos is given, stated in the following form:

$$\tilde{m}_1 + \tilde{m}_2 = 2\tilde{m}_3; \tilde{m}_j = m_j \exp(i\phi_j), \, j = 1,2,3. \quad (24)$$

A comparison of the general formulas (5) and (24) shows that:

$$p = +1, \Delta \chi 21 = 2\pi, \Delta \chi 31 = \pi, A_1 = 1, A_2 = 1, A_3 = 2. \quad (25)$$

This example is treated in Ref. [1], as an example that exclusively belongs to the inverted mass hierarchy: $m_3 < m_1 < m_2$. With this note, and bearing in mind that solutions for neutrino masses must be realistic, the mass sum rule for the Dirac neutrino should take the form of

$$m_2 = m_1 + 2m_3 \tag{26}$$

This will be confirmed in further consideration. Equation (24) is an equation with three unknowns $m_3, m_1, m_2$. Using the procedure as in the first example, this equation decomposes into two equations. The first equation reads:

$$m_1 + m_2 \cos(\alpha_{21}) = 2m_3 \cos(\alpha_{31}) \tag{27}$$

And the second one

$$m_2 \sin(\alpha_{21}) = 2m_3 \sin(\alpha_{31}) \tag{28}$$

As it is obvious, the second equation (28) is satisfied for all solutions related to the values:

$$\alpha_{21} = 0, \pm\pi, \pm 2\pi, \alpha_{31} = 0, \pm\pi, \pm 2\pi. \tag{29}$$

However, equations (27) and (28) have unique solutions that satisfy the conditions for inverted mass hierarchy and only in the case when the values are taken as solutions:

$$\alpha_{21} = \pi, \alpha_{31} = \pi. \tag{30}$$

Thus, taking into account solutions (30) it is obvious that equation (24) reduces to equation (26). And in this example, we have one equation with three unknowns: $m_1, m_2, m_3$, as already seen in the first example, so explicit numerical values for neutrino masses will be obtained by applying the same procedure. Squaring the left and right sides of this equation is crucial to define all the necessary equations to solve the unknown quantities. Thus, the following equations are obtained:

$$m_3^2 + m_1 m_3 - \frac{1}{4}\Delta m_{21}^2 = 0, \tag{31}$$

$$m_1^2 + 4m_1 m_3 + 3m_3^2 - \Delta m_{23}^2 = 0. \tag{32}$$

Relations (16) are also included in these equations, and by joining the equation (26), they form a complete set of equations of three unknown quantities. First we solve equations (31):

$$m_3 = -\frac{m_1}{2} \pm \frac{1}{2}\sqrt{m_1^2 + \Delta m_{21}^2} \qquad (33)$$

Then the solution for $m_3$ is inserted into the equation (32)

$$m_1^2 + 4m_1\left[-\frac{1}{2}m_1 \pm \frac{1}{2}\sqrt{m_1^2 + \Delta m_{21}^2}\right] + 3\left[-\frac{1}{2}m_1 \pm \frac{1}{2}\sqrt{m_1^2 + \Delta m_{21}^2}\right]^2 - \Delta m_{23}^2 = 0. \qquad (34)$$

By solving this equation, we find a solution for the numerical value for mass $m_1$:

$$m_1 = \frac{\Delta m_{23}^2 - 3/4\,\Delta m_{21}^2}{\sqrt{\Delta m_{23}^2 - 1/2\,\Delta m_{21}^2}} = \frac{U}{\sqrt{V}} = \frac{(0.002536 \pm 0.000034) - 3/4 \times (0.0000753 \pm 0.0000018)}{\sqrt{(0.002536 \pm 0.000034) - 1/2 \times (0.0000753 \pm 0.0000018)}} = \frac{\langle U \rangle}{\sqrt{\langle V \rangle}} + |\Delta m_1| = (0.0496068 \pm 0.000742)\,eV/c^2,$$

$$\langle m_1 \rangle = \frac{\langle U \rangle}{\sqrt{\langle V \rangle}} = \frac{0.002536 - 3/4 \times 0.0000753}{\sqrt{0.002536 - 1/2 \times 0.0000753}} \approx 0.04960687297\,eV/c^2, |\Delta m_1| = \left|\frac{\partial\left(\frac{\langle U \rangle}{\sqrt{\langle V \rangle}}\right)}{\partial \langle U \rangle}\right| |\Delta U| + \left|\frac{\partial\left(\frac{\langle U \rangle}{\sqrt{\langle V \rangle}}\right)}{\partial \langle V \rangle}\right| |\Delta V| = \frac{1}{\sqrt{\langle V \rangle}}|\Delta U| + \frac{1}{2}\left\langle\frac{U}{\sqrt{V}}\right\rangle\frac{1}{\langle V \rangle}|\Delta V| = \frac{1}{\sqrt{\langle V \rangle}}|\Delta U| + \frac{1}{2}\langle m_1 \rangle\frac{1}{\langle V \rangle}|\Delta V| \approx 0.000725\,eV/c,$$

$$|\Delta U| = (0.000034 + 3/4 \times 0.0000018)\,eV/c^2, |\Delta V| = (0.000034 + 1/2 \times 0.0000018)\,eV/c^2 \qquad (35)$$

Then, we arrive to the solution for the numerical value for $m_3$:

$$m_3 = -\frac{m_1}{2} + \frac{1}{2}\sqrt{m_1^2 + \Delta m_{21}^2} = -\frac{\langle m_1 \rangle}{2} + \frac{1}{2}\sqrt{\langle m_1 \rangle^2 + \langle \Delta m_{21}^2 \rangle} + \Delta\left(-\frac{m_1}{2} + \frac{1}{2}\sqrt{m_1^2 + \Delta m_{21}^2}\right),$$

$$\langle m_3 \rangle = -\frac{\langle m_1 \rangle}{2} + \frac{1}{2}\sqrt{\langle m_1 \rangle^2 + \langle \Delta m_{21}^2 \rangle} = 0.000376624306 \, eV/c^2,$$

$$\Delta\left(-\frac{m_1}{2} + \frac{1}{2}\sqrt{m_1^2 + \Delta m_{21}^2}\right) = \left|\frac{\Delta m_1}{2}\right| + \frac{1}{2}\left|\frac{\partial\left(\sqrt{\langle m_1 \rangle^2 + \langle \Delta m_{21}^2 \rangle}\right)}{\partial \langle m_1 \rangle}\right| |\Delta m_1| + \frac{1}{2}\left|\frac{\partial\left(\sqrt{\langle m_1 \rangle^2 + \langle \Delta m_{21}^2 \rangle}\right)}{\partial \langle \Delta m_{21}^2 \rangle}\right| \Delta\left(\Delta m_{21}^2\right) = \left|\frac{\Delta m_1}{2}\right| + \frac{1}{2}\frac{\langle m_1 \rangle}{\sqrt{\langle m_1 \rangle^2 + \langle \Delta m_{21}^2 \rangle}}|\Delta m_1| + \frac{1}{4}\frac{1}{\sqrt{\langle m_1 \rangle^2 + \langle \Delta m_{21}^2 \rangle}}\Delta\left(\Delta m_{21}^2\right) = 0.000729 \, eV/c^2, \quad (36)$$

$$m_3 \approx (0.000377 \pm 0.000729)\,eV/c^2, \left|\Delta\left(\Delta m_{21}^2\right)\right| = 0.0000018\,eV^2/c^4.$$

Finally, we find a value for $m_2$:

$$m_2 = m_1 + 2m_3 \approx (0.050361 \pm 0.002183)\,eV/c^2 \quad (37)$$

Thus, we obtained solutions for neutrino mass eigenstates as a function of the measured parameters: $\Delta m_{21}^2, \Delta m_{23}^2, \Delta m_{13}^2$ in case, the neutrinos would be the Dirac particles.

The cumulative value of the neutrino mass in this case is

$$\sum m_i = m_3 + m_2 + m_1 \approx (0.1004 \pm 0.0038)\,eV/c^2 \quad (38)$$

## 2.3. The third example. The anti-Dirac mass sum rule

The anti-Dirac mass sum rule we obtain by the changes in the notation of masses in the relation (24) in accordance with the normal hierarchy of masses

$$\tilde{m}_2 + \tilde{m}_3 = 2\tilde{m}_1; \tilde{m}_j = m_j \exp(i\phi_j), j = 1,2,3. \quad (39)$$

A comparison of the general formulas (5) and (39) shows that:

$$p = +1, \Delta\chi_{21} = \pi, \Delta\chi_{31} = \pi, A_1 = 2, A_2 = 1, A_3 = 2. \quad (40)$$

In contrast to relation (24), relation (39) belongs exclusively to the normal mass hierarchy: $m_1 < m_2 < m_3$. With this note, and bearing in mind that solutions for neutrino masses must be realistic, the mass sum rule for the anti-Dirac neutrino should take the form of

$$m_3 = m_2 + 2m_1 \tag{41}$$

This will be confirmed in further consideration.

Equation (39) is an equation with three unknowns $m_3, m_1, m_2$. Using the procedure as in previous examples, this equation decomposes into two equations. The first equation reads:

$$2m_1 - m_2 \cos(\alpha_{21}) = m_3 \cos(\alpha_{31}) \tag{42}$$

And the second one

$$-m_2 \sin(\alpha_{21}) = m_3 \sin(\alpha_{31}) \tag{43}$$

As it is obvious, the second equation (43) is satisfied for all solutions related to the values:

$$\alpha_{21} = 0, \pm\pi, \pm 2\pi, \alpha_{31} = 0, \pm\pi, \pm 2\pi. \tag{44}$$

However, equations (42) and (43) have unique solutions that satisfy the conditions for normal mass hierarchy and only in the case when the values are taken as solutions:

$$\alpha_{21} = \pi, \alpha_{31} = 0. \tag{45}$$

Thus, taking into account solutions (45) it is obvious that equation (39) reduces to equation (41). Squaring the left and right sides of this equation is crucial to define all the necessary equations to solve the unknown quantities. Thus, the following equations are obtained:

$$m_1^2 + m_1 m_2 - \frac{1}{4}\Delta m_{32}^2 = 0, \tag{46}$$

$$m_2^2 + 4m_1 m_2 + 3m_1^2 - \Delta m_{31}^2 = 0. \tag{47}$$

Relations (41) are also included in these equations, and they form a complete set of equations of three unknown quantities.

First we solve equations (46):

$$m_1 = -\frac{m_2}{2} \pm \frac{1}{2}\sqrt{m_2^2 + \Delta m_{32}^2} \qquad (48)$$

Then the solution for $m_1$ is inserted into the equation (47)

$$m_2^2 + 4m_2\left[-\frac{m_2}{2} \pm \frac{1}{2}\sqrt{m_2^2 + \Delta m_{32}^2}\right] + 3\left[-\frac{m_2}{2} \pm \frac{1}{2}\sqrt{m_2^2 + \Delta m_{32}^2}\right]^2 - \Delta m_{31}^2 = 0. \qquad (49)$$

By solving this equation, we find a solution for the numerical value for mass $m_2$:

$$m_2 = \frac{2\Delta m_{31}^2 - 3/2 \times \Delta m_{32}^2}{\sqrt{2\left(2\Delta m_{31}^2 - \Delta m_{32}^2\right)}} = \langle m_2 \rangle + |\Delta m_2| = \frac{\langle U \rangle}{\sqrt{\langle V \rangle}} + |\Delta m_2| = \frac{2 \times (0.0025283 \pm 0.0000348) - 3/2 \times (0.002453 \pm 0.000033)}{\sqrt{(4 \times (0.0025283 \pm 0.0000348) - 2 \times (0.002453 \pm 0.000033))}} \approx (0.019084 \pm 0.002027)\,eV/c^2,$$

$$\langle m_2 \rangle = \frac{\langle U \rangle}{\sqrt{\langle V \rangle}} = \frac{2 \times (0.0025283) - 3/2 \times (0.002453)}{\sqrt{(4 \times (0.0025283) - 2 \times (0.002453))}} = 0.019083733760\,eV/c^2, |\Delta m_2| = \frac{\partial\left(\frac{\langle U \rangle}{\sqrt{\langle V \rangle}}\right)}{\partial \langle U \rangle}|\Delta U| + \frac{\partial\left(\frac{\langle U \rangle}{\sqrt{\langle V \rangle}}\right)}{\partial \langle V \rangle}|\Delta V| = \frac{1}{\sqrt{\langle V \rangle}}|\Delta U| + \frac{1}{2}\frac{\langle U \rangle}{\sqrt{\langle V \rangle}}\frac{1}{V}|\Delta V| = \frac{1}{\sqrt{\langle V \rangle}}|\Delta U| + \frac{1}{2}\langle m_2 \rangle\frac{1}{V}|\Delta V|$$

$$= \frac{1}{\sqrt{(4 \times (0.0025283) - 2 \times (0.002453))}}|\Delta U| + \frac{1}{2}\langle m_2 \rangle\frac{1}{(4 \times (0.0025283) - 2 \times (0.002453))}|\Delta V| \qquad (50)$$

$$\approx 0.002027\,eV/c^2,$$

$$|\Delta U| = (2 \times 0.0000348 + 3/2 \times 0.000033)\,eV^2/c^4, |\Delta V| = (4 \times 0.0000348 + 2 \times 0.000033)\,eV^2/c^4.$$

Then, we arrive to the solution for the numerical value for $m_1$:

$$m_1 = -\frac{\langle m_2\rangle}{2} + \frac{1}{2}\sqrt{\langle m_2\rangle^2 + \langle\Delta m_{32}^2\rangle} = \langle m_1\rangle + \Delta m_1 \approx (0.016997 \pm 0.001494)eV/c^2, \langle m_1\rangle = -\frac{\langle m_2\rangle}{2} + \frac{1}{2}\sqrt{\langle m_2\rangle^2 + \langle\Delta m_{32}^2\rangle} = 0.016996731868 eV/c^2,$$

$$|\Delta m_1| = \frac{1}{2}|\Delta m_2| + \frac{1}{2}\left(\frac{\partial\left(\sqrt{\langle m_2\rangle^2 + \langle\Delta m_{32}^2\rangle}\right)}{\partial\langle m_2\rangle}|\Delta m_2| + \frac{\partial\left(\sqrt{\langle m_2\rangle^2 + \langle\Delta m_{32}^2\rangle}\right)}{\partial\langle\Delta m_{32}^2\rangle}\left|\Delta\left(\Delta m_{32}^2\right)\right|\right) = \frac{1}{2}|\Delta m_2| + \frac{1}{2}\frac{\langle m_2\rangle}{\sqrt{\langle m_2\rangle^2 + \langle\Delta m_{32}^2\rangle}}|\Delta m_2| + \frac{1}{4}\frac{1}{\sqrt{\langle m_2\rangle^2 + \langle\Delta m_{32}^2\rangle}}\left|\Delta\left(\Delta m_{32}^2\right)\right|,$$

$$|\Delta m_2| = 0.00202649 eV/c^2, \left|\Delta\left(\Delta m_{32}^2\right)\right| = 0.000033 eV^2/c^4.$$

(51)

Finally, we find a value for $m_2$:

$$m_3 = m_2 + 2m_1 = (0.053077 \pm 0.005014)eV/c^2 \tag{52}$$

Thus, we obtained solutions for neutrino mass eigenstates as a function of the measured parameters: $\Delta m_{21}^2, \Delta m_{32}^2, \Delta m_{31}^2$ in case, the neutrinos would be the Dirac particles.

The cumulative value of the neutrino mass in this case is

$$\sum m_i = m_3 + m_2 + m_1 = (0.08916 \pm 0.00853)eV/c^2 \tag{53}$$

## 2.4. The forth example. Group $A_4$, Seesaw Type Weinberg, Matrix $M_\nu$

This example is treated in [8] as a rule for the sum of neutrino masses

$$2\tilde{m}_2 + \tilde{m}_3 = \tilde{m}_1; \tilde{m}_j = m_j \exp(i\phi_j), j = 1,2,3. \tag{54}$$

A comparison of the general formulas (5) and (54) shows that:

$$p = +1, \Delta\chi_{21} = \pi, \Delta\chi_{31} = \pi, A_1 = 1, A_2 = 2, A_3 = 1. \tag{55}$$

In contrast to relation (24), relation (54) belongs exclusively to the normal mass hierarchy: $m_1 < m_2 < m_3$. With this note, and bearing in mind that solutions for neutrino masses must be realistic, the mass sum rule for the anti-Dirac neutrino should take the form of

$$m_3 = 2m_2 + m_1 \tag{56}$$

This will be confirmed in further consideration.

Equation (54) is an equation with three unknowns $m_3, m_1, m_2$. Using the procedure as in the previous cases, this equation decomposes into two equations. The first equation reads:

$$m_1 - 2m_2 \cos(\alpha_{21}) = m_3 \cos(\alpha_{31}) \tag{57}$$

And the second one

$$-2m_2 \sin(\alpha_{21}) = m_3 \sin(\alpha_{31}) \tag{58}$$

As it is obvious, the second equation (58) is satisfied for all solutions related to the values:

$$\alpha_{21} = 0, \pm\pi, \pm 2\pi, \alpha_{31} = 0, \pm\pi, \pm 2\pi. \tag{59}$$

However, equations (57) and (58) have unique solutions that satisfy the conditions for normal mass hierarchy and only in the case when the values are taken as solutions:

$$\alpha_{21} = \pi, \alpha_{31} = 0. \tag{60}$$

Thus, taking into account solutions (60) it is obvious that equation (57) reduces to equation (56). And in this example, we have one equation with three unknowns $m_1, m_2, m_3$, as already seen in the previous examples, so explicit numerical values for neutrino masses will be obtained by applying the same procedure. Squaring the left and right sides of this equation is crucial to define all the necessary equations to solve the unknown quantities. Thus, the following equations are obtained:

$$m_2^2 + m_1 m_2 - \frac{1}{4}\Delta m_{31}^2 = 0, \tag{61}$$

$$m_1^2 + 4m_1m_3 + 3m_2^2 - \Delta m_{32}^2 = 0. \tag{62}$$

Relations (16) are also included in these equations, and by joining the equation (56), they form a complete set of equations of three unknown quantities.

First we solve equation (61):

$$m_2 = -\frac{m_1}{2} \pm \frac{1}{2}\sqrt{m_1^2 + \Delta m_{31}^2} \tag{63}$$

Then the solution for $m_2$ is inserted into the equation (64)

$$m_1^2 + 4m_1\left[-\frac{m_1}{2} \pm \frac{1}{2}\sqrt{m_1^2 + \Delta m_{31}^2}\right] + 3\left[-\frac{m_1}{2} \pm \frac{1}{2}\sqrt{m_1^2 + \Delta m_{31}^2}\right]^2 - \Delta m_{32}^2 = 0. \tag{64}$$

By solving this equation, we find a solution for the numerical value for mass $m_1$:

$$m_1 = \frac{2\Delta m_{32}^2 - 3/2 \times \Delta m_{31}^2}{\sqrt{2(2\Delta m_{32}^2 - \Delta m_{31}^2)}} = \frac{2 \times (0.002453 \pm 0.000033) - 3/2 \times (0.0025283 \pm 0.0000348)}{\sqrt{(4 \times (0.002453 \pm 0.000033) - 2 \times (0.0025283 \pm 0.0000348))}} = \frac{\langle U \rangle}{\sqrt{\langle V \rangle}} + |\Delta m_1| = \langle m_1 \rangle + |\Delta m_1| \approx (0.016148 \pm 0.002051) eV/c^2,$$

$$\frac{\langle U \rangle}{\sqrt{\langle V \rangle}} = \frac{2 \times (0.002453) - 3/2 \times (0.0025283)}{\sqrt{(4 \times 0.002453) - 2 \times (0.0025283)}} = 0.016147905 eV/c^2,$$

$$|\Delta m_1| = \left|\frac{\partial\left(\frac{\langle U \rangle}{\sqrt{\langle V \rangle}}\right)}{\partial \langle U \rangle}\right||\Delta U| + \left|\frac{\partial\left(\frac{\langle U \rangle}{\sqrt{\langle V \rangle}}\right)}{\partial \langle V \rangle}\right||\Delta V| = \frac{1}{\sqrt{(4 \times 0.002453) - 2 \times (0.0025283)}}|\Delta U| + \frac{1}{2} \times \frac{2 \times (0.002453) - 3/2 \times (0.0025283)}{\sqrt{(4 \times 0.002453) - 2 \times (0.0025283)}}\frac{1}{(4 \times 0.002453) - 2 \times (0.0025283)}|\Delta V| \tag{65}$$

$$= \frac{1}{\sqrt{(4 \times 0.002453) - 2 \times (0.0025283)}}|\Delta U| + \frac{1}{2} \times \langle m_1 \rangle \frac{1}{(4 \times 0.002453) - 2 \times (0.0025283)}|\Delta V| = 0.002050225994 eV/c^2, |\Delta U| = [2 \times (00.000033) + 3/2 \times (0.0000348)]eV^2/c^4,$$

$$|\Delta V| = [(4 \times (0.000033) + 2 \times (0.0000348)]eV^2/c^4.$$

Then, we arrive to the solution for the numerical value for $m_2$:

$$m_2 = -\frac{m_1}{2} + \frac{1}{2}\sqrt{m_1^2 + \Delta m_{31}^2} = \langle m_2 \rangle + |\Delta m_2| \approx (0.018332 \pm 0.001392)eV/c^2, \langle m_2 \rangle = -\frac{\langle m_1 \rangle}{2} + \frac{1}{2}\sqrt{\langle m_1 \rangle^2 + \langle \Delta m_{31}^2 \rangle} = 0.018331798824 eV/c^2,$$

$$|\Delta m_2| = \frac{1}{2}|\Delta m_1| + \frac{1}{2}\left(\frac{\partial\left(\sqrt{\langle m_1 \rangle^2 + \langle \Delta m_{31}^2 \rangle}\right)}{\partial \langle m_1 \rangle}|\Delta m_1| + \frac{\partial\left(\sqrt{\langle m_1 \rangle^2 + \langle \Delta m_{31}^2 \rangle}\right)}{\partial \langle \Delta m_{31}^2 \rangle}\left|\Delta\left(\Delta m_{31}^2\right)\right|\right) = \frac{1}{2}|\Delta m_1| + \frac{1}{2}\frac{\langle m_1 \rangle}{\sqrt{\langle m_1 \rangle^2 + \langle \Delta m_{31}^2 \rangle}}|\Delta m_1| + \frac{1}{4}\frac{1}{\sqrt{\langle m_1 \rangle^2 + \langle \Delta m_{31}^2 \rangle}}\left|\Delta\left(\Delta m_{31}^2\right)\right|,$$

$$|\Delta m_1| = 0.002050225994 eV/c^2, \left|\Delta\left(\Delta m_{31}^2\right)\right| = 0.0000348 eV^2/c^4.$$

(66)

Finally, we find a value for $m_3$:

$$m_3 = 2m_2 + m_1 \approx (0.052812 \pm 0.004833)eV/c^2$$

(67)

The cumulative value of the neutrino mass in this case is

$$\sum m_i = m_3 + m_2 + m_1 \approx (0.087292 \pm 0.007275)eV/c^2$$

(68)

## 2.5. *Discussion of the obtained results*

In this theoretical consideration, two examples of the mass sum rules that satisfy the condition of the inverted neutrino mass hierarchy are analyzed. The first example is related to the Weinberg operator (6) and the second to the Dirac neutrinos (24) [1].While Examples 3 and 4 belong to the normal neutrino mass hierarchy. The calculated theoretical results (21,35,36,37) and (50,51,52,53,65,66,67) are compared with the results obtained by processing data from experimental measurements in the form of a graphical presentation of possible neutrino masses given in [9], and they are shown in Figure 1.If we take into account the above limitation in cosmology for the sum of masses $\sum m_i < 0.12 eV/c^2$, the following can be concluded:1. The first example does not meet the condition for the limit value in cosmology.2. The second example satisfies the condition for the limit value in cosmology. However, the calculation of individual values for neutrino masses is not in accordance with the definition for Dirac particles. So this example must be omitted.3. The third example satisfies the boundary condition in cosmology.4. The fourth example satisfies the boundary condition in cosmology.It is interesting to note that the calculated values for neutrino masses in examples 1, 3 and 4 are consistent with the values shown in Figure 1. While this could not be concluded separately for the value of $m_3$ in the second example.Therefore, the result given in formula (36) stands out. Namely, the theoretically calculated absolute error is greater than the average value. Such an absurd result can only mean that such a physical quantity could not exist in nature.

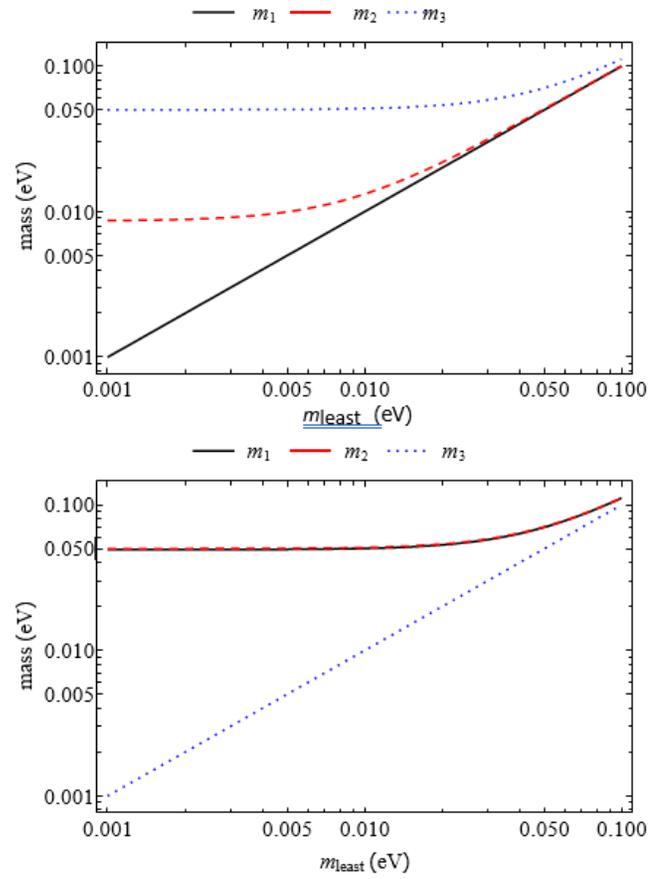

**Figure 1: Current best-fit values of the neutrino masses $m_1, m_2, m_3$ as a function of the lightest neutrino mass, for the normal mass ordering (top) and inverted mass ordering**.(bottom) [9] .

So, based on that result, it could be explained why the graph $m_3$ in Figure 1 (Inverted mass ordering) cannot achieve the value of $m_{least} = m_3 \approx 0.000377 \; eV$.

### 3. Application of the mathematical model to determine the Dirac CP violation phase

Here, we will apply the formula for calculating the Dirac CP violation phase, to the data processed, based on experimental measurements given in [3].

In the appendix of this paper, the procedure for reaching the equation of motion of neutrinos is given. A general equation of neutrino motion is derived, in which the magnitude of the Dirac CP violation phase is unknown $\delta_{CP}[0,2\pi)$, which reads (A28);

$$(2W\cos\delta_{CP} - V\sin\delta_{CP}) = 0 \tag{69}$$

Writing the same equation (69) for the published data for the neutrino parameters in [3] there is a separate Dirac CP violating phase for each hierarchy.

*Normal ordering*

In this case, the equation for the Dirac CP violation phase is:

$$(2W\cos\delta_{CP} - V\sin\delta_{CP}) = 0 \tag{70}$$

$$\delta CP(NO) = arctg\left(\frac{2W}{V}\right) = arctg\left[2\frac{\sin^2\left(\pi\frac{\Delta m_{31}^2}{\Delta m_{21}^2}\right)}{\sin\left(2\pi\frac{\Delta m_{31}^2}{\Delta m_{21}^2}\right)}\right] = arctg\left[2\frac{\sin^2\left(\pi\frac{0.0025283 \pm 0.0000348}{0.0000753 \pm 0.0000018}\right)}{\sin\left(2\pi\frac{0.0025283 \pm 0.0000348}{0.0000753 \pm 0.0000018}\right)}\right] = arctg\left[tg\left(\pi\frac{\Delta m_{31}^2}{\Delta m_{21}^2}\right)\right] = arctg\left[tg\left(\pi\frac{0.0025283 \pm 0.0000348}{0.0000753 \pm 0.0000018}\right)\right]$$

$$\approx arctg\left[tg\left(\pi\frac{0.0025283}{0.0000753}\right)\right] + |\Delta(\delta CP(NO))| = \langle\delta CP(NO)\rangle + |\Delta(\delta CP(NO))|. \tag{71}$$

Where

$$\langle W \rangle = \sin^2\left(\pi \frac{\langle \Delta m_{31}^2 \rangle}{\langle \Delta m_{21}^2 \rangle}\right) = \sin^2\left(\pi \frac{\langle \Delta m_{32}^2 \rangle}{\langle \Delta m_{21}^2 \rangle}\right) = 0.9435455445172244, \langle V \rangle = \sin\left(2\pi \frac{\langle \Delta m_{31}^2 \rangle}{\langle \Delta m_{21}^2 \rangle}\right) = \sin\left(2\pi \frac{\langle \Delta m_{32}^2 \rangle}{\langle \Delta m_{21}^2 \rangle}\right) = -0.461594410446723,$$

$$\langle \Delta m_{21}^2 \rangle = 0.0000753 eV^2, \langle \Delta m_{31}^2 \rangle = 0.0025283 eV^2, \langle \Delta m_{32}^2 \rangle = 0.002453 eV^2.$$

(72)

The average value is:

$$\langle \delta_{CP}(NO) \rangle = arctg\left[tg\left(\pi \frac{0.0025283}{0.0000753}\right)\right] \approx -76.255^0 = +283.745^0, \langle \delta_{CP}(NO) \rangle = 180^0 \times \frac{\langle \Delta m_{31}^2 \rangle}{\langle \Delta m_{21}^2 \rangle} = 180^0 \times \frac{0.0025283}{0.0000753} \approx +283.745^0.$$

(73)

and the error of the indirectly measured physical quantity amounts to

$$|\Delta(\delta_{CP}(NO))| = 180^0 \times \frac{1}{(0.0000753)} \times 0.0000348 + 180^0 \times \frac{(0.0025283)}{(0.0000753)^2} \times 0.0000018 = |227.659^0|, |\Delta(\Delta_{31}^2)| = 0.0000348 eV^2, |\Delta(\Delta_{21}^2)| = 0.0000018 eV^2$$

(74)

So it is

$$\delta_{CP}(NO) = \langle \delta_{CP}(NO) \rangle + |\Delta(\delta_{CP}(NO))| \approx 283.745^0 \pm 227.659^0$$

(75)

*Inverted ordering*

In this case will be:

$$\delta_{CP}(IO) = arctg\left(\frac{2W}{V}\right) = arctg\left[2 \frac{\sin^2\left(\pi \frac{\Delta m_{23}^2}{\Delta m_{21}^2}\right)}{\sin\left(2\pi \frac{\Delta m_{23}^2}{\Delta m_{21}^2}\right)}\right] = arctg\left[2 \frac{\sin^2\left(\pi \frac{0.002536 \pm 0.000034}{0.0000753 \pm 0.0000018}\right)}{\sin\left(2\pi \frac{0.002536 \pm 0.000034}{0.0000753 \pm 0.0000018}\right)}\right] = arctg\left[tg\left(\pi \frac{\Delta m_{23}^2}{\Delta m_{21}^2}\right)\right] = arctg\left[tg\left(\pi \frac{0.002536 \pm 0.000034}{0.0000753 \pm 0.0000018}\right)\right] \approx arctg\left[tg\left(\pi \frac{0.002536}{0.0000753}\right)\right] + |\Delta(\delta_{CP}(IO))| = \langle \delta_{CP}(IO) \rangle + |\Delta(\delta_{CP}(IO))|$$

(76)

Where

$$\langle W \rangle = \sin^2\left(\pi \frac{\langle \Delta m_{13}^2 \rangle}{\langle \Delta m_{21}^2 \rangle}\right) = \sin^2\left(\pi \frac{\langle \Delta m_{23}^2 \rangle}{\langle \Delta m_{21}^2 \rangle}\right) = 0.716807612, \langle V \rangle = \sin\left(2\pi \frac{\langle \Delta m_{13}^2 \rangle}{\langle \Delta m_{21}^2 \rangle}\right) = \sin\left(2\pi \frac{\langle \Delta m_{23}^2 \rangle}{\langle \Delta m_{21}^2 \rangle}\right) = -0.901098128,$$

$$\langle \Delta m_{21}^2 \rangle = 0.0000753 eV^2, \langle \Delta m_{13}^2 \rangle = 0.0024607 eV^2, \langle \Delta m_{23}^2 \rangle = 0.002536 eV^2.$$

(77)

The average value is:

$$\langle \delta_{CP}(IO) \rangle = arctg\left[tg\left(\pi \frac{\Delta m_{23}^2}{\Delta m_{21}^2}\right)\right] \approx arctg\left[tg\left(\pi \frac{0.002536}{0.0000753}\right)\right] \approx -57.848^0 = +302.152^0,$$

$$\langle \delta_{CP}(IO) \rangle = 180^0 \times \frac{\Delta m_{23}^2}{\Delta m_{21}^2} = 180^0 \times \frac{0.002536}{0.0000753} \approx +302.152^0.$$

(78)

and the error of the indirectly measured physical quantity amounts to

$$\left|\Delta(\delta_{CP}(IO))\right| = \Delta\left(180^0 \times \frac{\Delta m_{23}^2}{\Delta m_{21}^2}\right) = \Delta\left[180^0 \times \frac{\langle \Delta m_{23}^2 \rangle \pm \Delta(\Delta m_{23}^2)}{\langle \Delta m_{21}^2 \rangle \pm \Delta(\Delta m_{21}^2)}\right] = \Delta\left(180^0 \times \frac{0.002536 \pm 0.000034}{0.0000753 \pm 0.0000018}\right) = 180^0 \times \frac{1}{(0.0000753)} \times \left|\Delta(\Delta m_{23}^2)\right| + 180^0 \times \frac{0.002536}{(0.0000753)^2} \times \left|\Delta(\Delta m_{21}^2)\right|$$

$$\approx \left|226.187^0\right|, \left|\Delta(\Delta m_{23}^2)\right| = 0.000034 eV^2, \left|\Delta(\Delta m_{21}^2)\right| = 0.0000018 eV^2, \langle \Delta m_{23}^2 \rangle = 0.002536 eV^2, \langle \Delta m_{21}^2 \rangle = 0.0000753 eV^2.$$

(79)

So it is

$$\delta_{CP}(IO) = \langle \delta_{CP}(IO) \rangle + \left|\Delta(\delta_{CP}(IO))\right| \approx 302.152^0 \pm 226.187^0$$

(80)

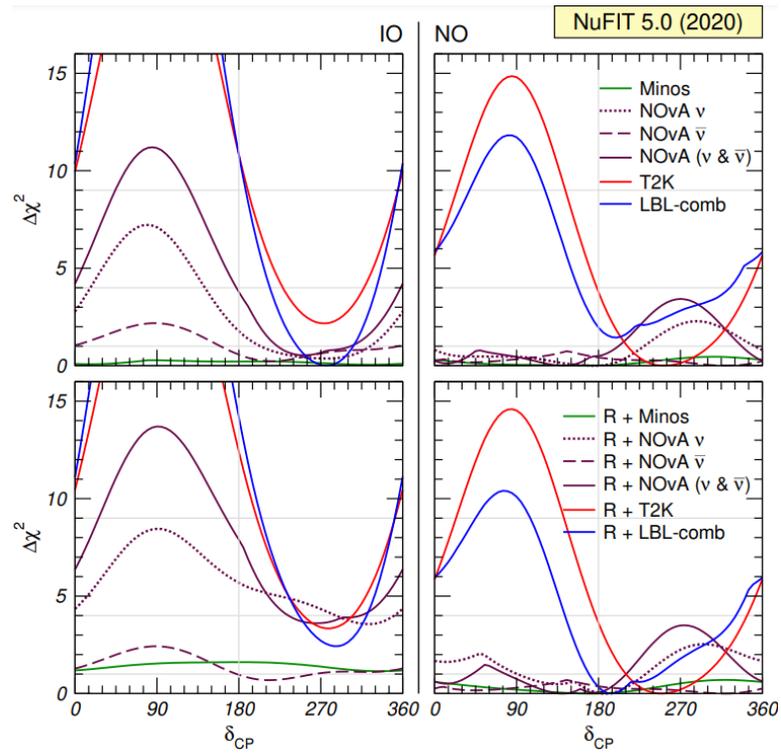

**Figure 2: Determination of $\delta_{CP}$ [10]**

The calculated values of (73) and (78) can be found in the graphs of Figure 2.

## 4. Prediction of the effective Majorana mass

There are opinions that if the process of neutrinoless double beta decay were confirmed by measuring the effective value of the neutrino Majorana mass, then absolute values for individual neutrino mass values could be obtained [11] [12]. To establish connections between the polar phase $\phi_j; j=1,2,3$, that are in fact Majorana phases, and phases $\alpha_{j1}; j=2,3$ which figure in the formula for the effective values of the Majorana mass, it is necessary to include the definition of both parametrizations: PDG parametrization and Symmetry parametrization.

*PDG parametrization: Jarlskog invariant*

The first step in this procedure is to define the Jarlskog invariant in both parameterizations, PDG parametrization, and Symmetry parametrization, as given in [1]. Inserting PDG parametrization

$$U_{PMNS}^{PDG} = \begin{pmatrix} c_{12}c_{13} & s_{12}c_{13} & s_{13}e^{-i\delta_{CP}} \\ -s_{12}c_{23}-c_{12}s_{23}s_{13}e^{i\delta_{CP}} & c_{12}c_{23}-s_{12}s_{13}s_{23}e^{i\delta_{CP}} & s_{23}c_{13} \\ s_{12}s_{23}-c_{12}c_{23}s_{13}e^{i\delta_{CP}} & -c_{12}s_{23}-s_{12}s_{13}c_{23}e^{i\delta_{CP}} & c_{23}c_{13} \end{pmatrix} \begin{pmatrix} 1 & 0 & 0 \\ 0 & e^{i\frac{\alpha_{21}}{2}} & 0 \\ 0 & 0 & e^{i\frac{\alpha_{31}}{2}} \end{pmatrix}$$

(81)

the Majorana effective mass is defined by the formula [1,4,5]

$$m_{ee}^{PDG} = \left| m_1 c_{12}^2 c_{13}^2 + m_2 s_{12}^2 c_{13}^2 \exp[i\alpha_{21}] + m_3 s_{13}^2 \exp[i(\alpha_{31}-2\delta_{CP})] \right|$$

(82)

It should be noted here that in formula (81) phases $\alpha_{21}, \alpha_{31}$ figure, which are already determined by formulas (45, 60), and they would be sufficient to calculate the effective mass (82). However, if they are the real ones, another parametrization is being introduced in the form of the products of the following partial matrices [1]:

$$U_{PMNS}^{Sym} = \omega_{12}(\theta_{12},\phi_{12})\omega_{13}(\theta_{13},\phi_{13})\omega_{23}(\theta_{23},\phi_{23}) = \begin{pmatrix} c_{12} & s_{12}\exp(-i\phi_{12}) & 0 \\ -s_{12}\exp(i\phi_{12}) & c_{12} & 0 \\ 0 & 0 & 1 \end{pmatrix} \begin{pmatrix} c_{13} & 0 & s_{13}\exp(-i\phi_{13}) \\ 0 & 1 & 0 \\ -s_{13}\exp(i\phi_{13}) & 0 & c_{13} \end{pmatrix} \begin{pmatrix} 1 & 0 & 0 \\ 0 & c_{23} & s_{23}\exp(-i\phi_{23}) \\ 0 & -s_{23}\exp(i\phi_{23}) & c_{23} \end{pmatrix}$$

(83)

In this parameterization, Jarlskog invariant is present

$$J_{CP}^{Sym} = s_{12}c_{12}s_{23}c_{23}s_{13}c_{13}^2 \sin(\phi_{13} - \phi_{12} - \phi_{23}) \tag{84}$$

In PDG parametrization, where a matrix is applied $U_{PMNS}^{PDG}$, Jarlskog invariant is present:

$$J_{CP}^{PDG} = s_{12}c_{12}s_{23}c_{23}s_{13}c_{13}^2 \sin\delta_{CP} = J_{CP}^{max} \sin\delta_{CP} = s_{12}c_{12}s_{23}c_{23}s_{13}c_{13}^2 \sin\left[180^0 \times \frac{\Delta m_{31}^2}{\Delta m_{21}^2}\right] = J_{CP}^{max} \sin\left[180^0 \times \frac{\Delta m_{31}^2}{\Delta m_{21}^2}\right], \delta_{CP} = \left[180^0 \times \frac{\Delta m_{31}^2}{\Delta m_{21}^2}\right]. \tag{85}$$

And both formulas (84) and (85) are by definition equal to each other

$$J_{CP}^{Sym} = J_{CP}^{PDG} \tag{86}$$

Thus, a connection was established between the Dirac CP violation phase $\delta_{CP}$ and phases $\phi_{12}, \phi_{13}, \phi_{23}$ :

$$\sin(\phi_{13} - \phi_{12} - \phi_{23}) = \sin\delta_{CP} \tag{87}$$

We can recall that complex neutrino masses are displayed in polar form $\tilde{m}_j = m_j \exp(i\phi_j), j = 1,2,3$, where $m_j > 0$ are the physical neutrino mass eigenvalues, and $\phi_j \in [0.2\pi)$ are the Majorana phases. From equality (84,85) follows the equality of effective values of the Majorana mass

$$\left[m_{ee}^{Sym}\right] = \left[m_{ee}^{PDG}\right] \tag{88}$$

Where

$$\left[m_{ee}^{Sym}\right] = \left|m_1 c_{12}^2 c_{13}^2 + m_2 s_{12}^2 c_{13}^2 \exp(i2\phi_{12}) + m_3 s_{13}^2 \exp(i2\phi_{13})\right|,$$
$$\left[m_{ee}^{PDG}\right] = \left|m_1 c_{12}^2 c_{13}^2 + m_2 s_{12}^2 c_{13}^2 \exp(i\alpha_{21}) + m_3 s_{13}^2 \exp[i(\alpha_{31} - 2\delta_{CP})]\right|. \tag{89}$$

Based on (84), (85), (88) and (89) connections were established between the Dirac CP phase $\delta_{CP}$ and phases $\phi_{12}, \phi_{13}, \phi_{23}, \alpha_{21}, \alpha_{31}$, which is given in Ref. [1] via matrix transformations:

$$\begin{pmatrix} \phi_{12} \\ \phi_{13} \\ \phi_{23} \end{pmatrix} = \begin{pmatrix} 0 & 1/2 & 0 \\ -1 & 0 & 1/2 \\ -2 & -1/2 & 1/2 \end{pmatrix} \begin{pmatrix} \delta_{CP} \\ \alpha_{21} \\ \alpha_{31} \end{pmatrix} \tag{90}$$

$$\begin{pmatrix} \delta_{CP} \\ \alpha_{21} \\ \alpha_{31} \end{pmatrix} = \begin{pmatrix} -1 & 1 & -1 \\ 2 & 0 & 0 \\ -2 & 4 & -2 \end{pmatrix} \begin{pmatrix} \phi_{12} \\ \phi_{13} \\ \phi_{23} \end{pmatrix} \tag{91}$$

We use the matrix transformation (90) to determine the parameters $\phi_{12}, \phi_{13}, \phi_{23}$ in function of Majorana phases (45, 60) and $\delta_{CP}$

$$\begin{pmatrix} \phi_{12} \\ \phi_{13} \\ \phi_{23} \end{pmatrix} = \begin{pmatrix} 0 & 1/2 & 0 \\ -1 & 0 & 1/2 \\ -2 & -1/2 & 1/2 \end{pmatrix} \begin{pmatrix} \delta_{CP} \\ \pi \\ 0 \end{pmatrix} \tag{92}$$

From here we find:

$$\phi_{12} = \frac{\pi}{2}, \phi_{13} = -\delta_{CP}, \phi_{23} = -2\delta_{CP} - \frac{1}{2}\pi. \tag{93}$$

And then we use these parameters (93) to calculate the Jarlskog invariant:

$$J_{CP}^{Sym} = s_{12}c_{12}s_{23}c_{23}s_{13}c_{13}^2 \sin(\phi_{13} - \phi_{23} - \phi_{12}) = J_{CP}^{max} \sin\left(-\delta_{CP} + 2\delta_{CP} + \frac{1}{2}\pi - \frac{1}{2}\pi\right) = J_{CP}^{max} \sin\delta_{CP} \tag{94}$$

And on the other hand by applying matrix transformation (91)

$$\begin{pmatrix} \delta_{CP} \\ \alpha_{21} \\ \alpha_{31} \end{pmatrix} = \begin{pmatrix} -1 & 1 & -1 \\ 2 & 0 & 0 \\ -2 & 4 & -2 \end{pmatrix} \begin{pmatrix} \pi/2 \\ -\delta_{CP} \\ -2\delta_{CP} - 1/2\pi \end{pmatrix} \tag{95}$$

it is :

$$\delta_{CP} = -\phi_{12} + \phi_{13} - \phi_{23}, \alpha_{21} = \pi, \alpha_{31} = 0. \tag{96}$$

The results of the third and fourth examples are presented here using the following calculated data:

*The third example*

$$\begin{aligned} m_1 &= (0.016997 \pm 0.001494)\,eV/c^2 \approx (0.016997)\,eV/c^2, \\ m_2 &= (0.019084 \pm 0.002027)\,eV/c^2 \approx (0.019084)\,eV/c^2, \\ m_3 &= (0.053077 \pm 0.005013)\,eV/c^2 \approx (0.053077)\,eV/c^2. \end{aligned} \tag{97}$$

*The fourth example*

$$\begin{aligned} m_1 &= (0.016148 \pm 0.002051)\,eV/c^2 \approx (0.016148)\,eV/c^2, \\ m_2 &= (0.018332 \pm 0.001392)\,eV/c^2 \approx (0.018332)\,eV/c^2, \\ m_3 &= (0.052812 \pm 0.004833)\,eV/c^2 \approx (0.052812)\,eV/c^2. \end{aligned} \tag{98}$$

Common data are:

$$\delta_{CP} \approx 283.745^0, c_{12}^2 \approx 0.693, c_{13}^2 \approx 0.978, s_{13}^2 \approx 0.022, \alpha_{21} = \pi, \alpha_{31} = 0. \tag{99}$$

*Numerous values for the effective Majorana neutrino mass*

*The third example*

$$\begin{aligned} \left|m_{ee}^{PDG}\right| &= \left|m_{ee}^{Sym}\right| = \left|m_1 c_{12}^2 c_{13}^2 + m_2 s_{12}^2 c_{13}^2 \exp(i2\phi_{12}) + m_3 s_{13}^2 \exp(i2\phi_{13})\right| \approx \left|m_1 c_{12}^2 c_{13}^2 + m_2 s_{12}^2 c_{13}^2 [\exp(i\alpha_{21})] + m_3 s_{13}^2 \exp[i(\alpha_{31} - 2\delta_{CP})]\right| \\ &\approx \left|m_1 c_{12}^2 c_{13}^2 - m_2 s_{12}^2 c_{13}^2 + m_3 s_{13}^2 \langle \cos 2\delta_{CP} \rangle - i m_3 s_{13}^2 \langle \sin 2\delta_{CP} \rangle\right| = 0.0047835\,eV/c^2 \approx 0.0048\,eV/c^2 \end{aligned} \tag{100}$$

*The fourth example*

$$\left||m_{ee}^{PDG}\right||=\left||m_{ee}^{Sym}\right||=\left|m_1C_{12}^2C_{13}^2+m_2S_{12}^2C_{13}^2\exp(i2\phi_{12})+m_3S_{13}^2\exp(i2\phi_{13})\right|\approx\left|m_1C_{12}^2C_{13}^2+m_2S_{12}^2C_{13}^2[\exp(i\alpha_{21})]+m_3S_{13}^2\exp[i(\alpha_{31}-2\delta_{CP})]\right|$$
$$\approx\left|m_1C_{12}^2C_{13}^2-m_2S_{12}^2C_{13}^2+m_3S_{13}^2\langle\cos 2\delta_{CP}\rangle-im_3S_{13}^2\langle\sin 2\delta_{CP}\rangle\right|=0.004442\,eV/c^2\approx 0.004\,eV/c^2$$

(101)

The calculated values (100) and (101) can be seen in Figure 3.

On the other hand, we have a process where observations of double beta decay without emission of neutrinos are researched, which is described by the following nuclear reaction [5] [11] [12] [13] [14][15]:

$$^A_Z X_N \rightarrow ^A_{Z+2} X_{N-2} + 2e^-$$

(102)

If experiments were to detect this reaction, it would be a sign that the neutrinos are Majorana particles. The $0\nu\beta\beta$ transition rate is expressed in terms of the following quantities, which are estimated to be by value in the next intervals [5]:

$$|m_{ee}|=\frac{m_e}{|M^{0\nu}|}\sqrt{\frac{1}{T_{1/2}g_A G^{0\nu}}}\approx(\min=0.0015,\max=0.004)\,eV\,(NO)$$

(103)

$$|m_{ee}|=\frac{m_e}{|M^{0\nu}|}\sqrt{\frac{1}{T_{1/2}g_A G^{0\nu}}}\approx(\min=0.017,\max=0.045)\,eV\,(IO)$$

(104)

Where: $m_e$ is the electron mass, $M^{0\nu}$ is nuclear matrix element which describes the actual decays in the nuclear environment, $T_{1/2}$ is the half-life of neutrinoless double-beta decay, $g_A$ -is the axial-vector weak coupling, $G^{0\nu}$ is a phase-space factor which determines how many electrons have the right energies and momenta to participate in the process.

We expect that in experiments [5], if a signal in the range (103) were detected then the hierarchy of neutrinos would be normal, while the appearance of the signal in the range (104) would be a confirmation that the neutrinos belong to the inverted mass hierarchy.

If we take into account the above limitation in cosmology for the sum of masses $\sum m_i < 0.12 eV/c^2$, the following can be concluded:

1. The first example does not meet the condition for the limit value in cosmology.

2. The second example satisfies the condition for the limit value in cosmology. However, the calculation of individual values for neutrino masses is not in accordance with the definition for Dirac particles. So this example must be omitted.

3. The third example satisfies the boundary condition in cosmology.

4. The fourth example satisfies the boundary condition in cosmology.

It is interesting to note that the calculated values for neutrino masses in Examples 1, 3 and 4 are consistent with the values shown in Figure 1. While this could not be concluded separately for the value of $m_3$ in the second example.

The calculated cumulative values for neutrino masses (The first example) $\sum m_i \approx 0.138 eV/c^2$ are higher than the upper limit projected in cosmology $\sum m_i < 0.12 eV/c^2$, so the possibility of the existence of an inverse hierarchy of neutrino masses in nature could be ruled out.

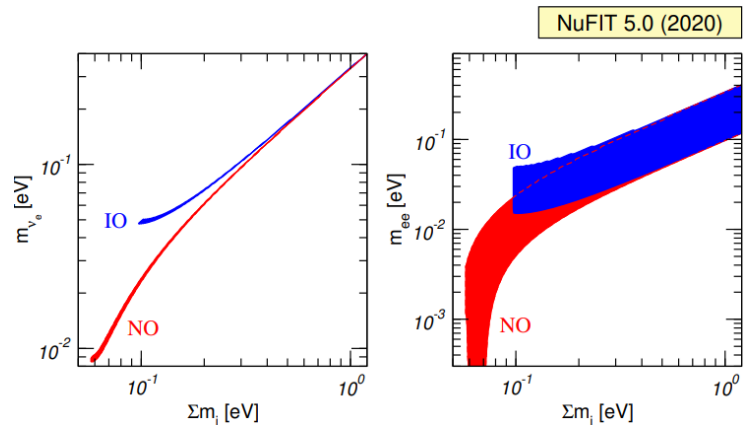

**Figure 3: Neutrino mass scale observables [10].**

Since Majorana neutrino phases are completely determined (45, 60), the matrix (81) can be written in the explicit form.

$$U_{PMNS}^{PDG} = \begin{pmatrix} c_{12}c_{13} & s_{12}c_{13} & s_{13}e^{-i\delta_{CP}} \\ -s_{12}c_{23}-c_{12}s_{23}s_{13}e^{i\delta_{CP}} & c_{12}c_{23}-s_{12}s_{13}s_{23}e^{i\delta_{CP}} & s_{23}c_{13} \\ s_{12}s_{23}-c_{12}c_{23}s_{13}e^{i\delta_{CP}} & -c_{12}s_{23}-s_{12}s_{13}c_{23}e^{i\delta_{CP}} & c_{23}c_{13} \end{pmatrix} \begin{pmatrix} 1 & 0 & 0 \\ 0 & e^{i\frac{\alpha_{21}}{2}} & 0 \\ 0 & 0 & e^{i\frac{\alpha_{31}}{2}} \end{pmatrix} = \begin{pmatrix} c_{12}c_{13} & s_{12}c_{13} & s_{13}e^{-i\delta_{CP}} \\ -s_{12}c_{23}-c_{12}s_{23}s_{13}e^{i\delta_{CP}} & c_{12}c_{23}-s_{12}s_{13}s_{23}e^{i\delta_{CP}} & s_{23}c_{13} \\ s_{12}s_{23}-c_{12}c_{23}s_{13}e^{i\delta_{CP}} & -c_{12}s_{23}-s_{12}s_{13}c_{23}e^{i\delta_{CP}} & c_{23}c_{13} \end{pmatrix} \begin{pmatrix} 1 & 0 & 0 \\ 0 & e^{i\frac{\pi}{2}} & 0 \\ 0 & 0 & e^{i0} \end{pmatrix}$$

$$= \begin{pmatrix} c_{12}c_{13} & s_{12}c_{13} & s_{13}e^{-i\delta_{CP}} \\ -s_{12}c_{23}-c_{12}s_{23}s_{13}e^{i\delta_{CP}} & c_{12}c_{23}-s_{12}s_{13}s_{23}e^{i\delta_{CP}} & s_{23}c_{13} \\ s_{12}s_{23}-c_{12}c_{23}s_{13}e^{i\delta_{CP}} & -c_{12}s_{23}-s_{12}s_{13}c_{23}e^{i\delta_{CP}} & c_{23}c_{13} \end{pmatrix} \begin{pmatrix} 1 & 0 & 0 \\ 0 & i & 0 \\ 0 & 0 & 1 \end{pmatrix}$$

## 5. Conclusion

In cosmology, it has been established that the upper limit for the sum of neutrino masses is $\sum m_i < 0.12 eV/c^2$. Analysis of the results for the four selected examples shows that the third and fourth examples, among other features, meet this criterion. This could also mean that a normal mass hierarchy has an advantage over an inverse mass hierarchy. In this paper, we developed a procedure for calculating the physical characteristics of neutrinos, such as: The absolute mass of neutrinos, the values of Majorana phases, the Dirac CP violating phase, the value of the effective Majorana neutrino mass, the Majorana matrix We have shown that neutrinos could be Majorana particles, i.e. a neutrino is its own antiparticle..

The Jarlskog invariant has been published in papers [16] [18] in general form by formula $J_{CP}^{Sym} = s_{12}c_{12}s_{23}c_{23}s_{13}c_{13}^2 \sin\delta_{CP} = J_{CP}^{\max}\sin\delta_{CP}$. Due to the ignorance of the explicit numerical value for the Dirac CP phase in neutrino physics, it is represented by computer simulations with graphs as shown in Figure 4.

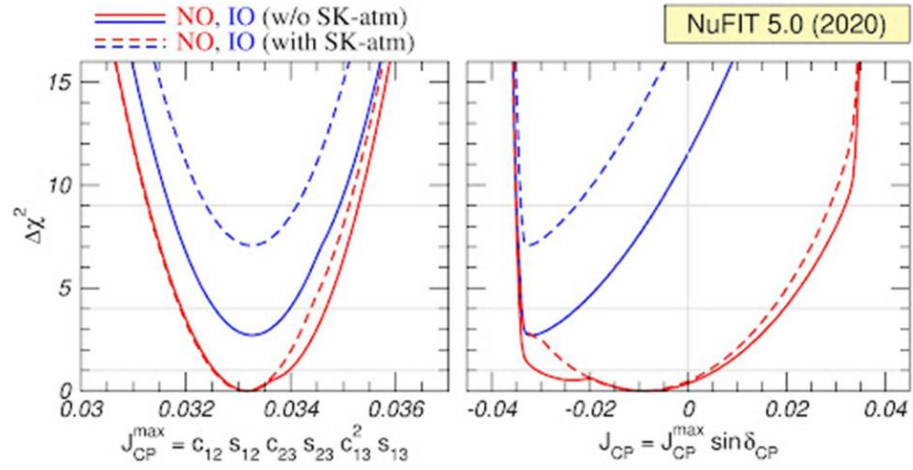

**Figure 4: CP violation: Jarlskog invariant [10]**

By deriving the neutrino equations (A4) and (A16) in which the Dirac CP phase is an unknown quantity and finding its explicit value, the Jarlskog invariant obtains a final version:

*Normal ordering*

$$J_{CP}^{PDG}(NO) = \left(s_{12}c_{12}s_{23}c_{23}s_{13}c_{13}^2\right)_{NO} \sin\delta_{CP(NO)} = J_{CP}^{max}(NO) \sin\delta_{CP(NO)} = \left(s_{12}c_{12}s_{23}c_{23}s_{13}c_{13}^2\right)_{NO} \sin\left[180^0 \times \frac{\Delta m_{31}^2}{\Delta m_{21}^2}\right]$$

$$= J_{CP}^{max}(NO) \sin\left[180^0 \times \frac{\Delta m_{31}^2}{\Delta m_{21}^2}\right], \delta_{CP(NO)} = \left[180^0 \times \frac{\Delta m_{31}^2}{\Delta m_{21}^2}\right].$$

*Inverted ordering*

$$J_{CP}^{PDG}(IO) = \left(s_{12}c_{12}s_{23}c_{23}s_{13}c_{13}^2\right)_{IO} \sin\delta_{CP(IO)} = J_{CP}^{max}(IO) \sin\delta_{CP(IO)} = \left(s_{12}c_{12}s_{23}c_{23}s_{13}c_{13}^2\right)_{IO} \sin\left[180^0 \times \frac{\Delta m_{23}^2}{\Delta m_{21}^2}\right]$$

$$= J_{CP}^{max}(IO) \sin\left[180^0 \times \frac{\Delta m_{23}^2}{\Delta m_{21}^2}\right], \delta_{CP(IO)} = \left[180^0 \times \frac{\Delta m_{23}^2}{\Delta m_{21}^2}\right].$$

The effective Majorana mass is defined through the process of atomic nucleus decay using Feynman diagram, which is the subject of experimental measurements [5] [7] [11]. Thus, we have two ways of obtaining the effective Majorana mass: one is defined by formulas (100,101), and the second by formulas (103, 104). On one hand, we have that in formulas (100,101) a connection is established between neutrino masses, Majorana phases and the Dirac CP violation phase, and on the other hand, we have formulas (103, 104) in which the half- life of $0\nu\beta\beta$ decay in the nuclear environment is present. The significance in the experimental measurement of the determining the upper limit for $T_{1/2}$ is that it can be converted into an upper limit on the effective Majorana neutrino mass under the assumption that the decay is dominated by the exchange of light Majorana mass [11] [12].

The resolution on the nature of neutrinos is based on comparing the numerical values for the effective Majorana neutrino mass (100) and (101) with the values on the graph in Figure 3. Where we can see, that the Majorana neutrinos have an advantage over the Dirac neutrinos.

An experiment, where the emission of two electrons without the emission of neutrinos would be detected, has not yet taken place. And when that event occurs, then it is unequivocal to follow that Majorana neutrinos are particles belonging to the normal mass hierarchy only if the mutually consistent values are defined by formulas (100,101) and (103).

## A. Appendix

### A.1 Application of the $U_{PMNS}^{PDG}$ mixing matrix

In the processes known as neutrino flavor oscillations, the Dirac CP violation phase $\delta_{CP}$ is unequivocally singlet out as the cause of those oscillations in the propagation of the neutrino beam through the physical vacuum. For that reason, there arises the question of writing the equation of motion in which $\delta_{CP}$ would appear as an unknown quantity. On the basis of that equation, it would be possible to determine that unknown quantity. So far, there appears to be only one way to derive equations of motion for a neutrino beam, and it is related to the use of the equations of the neutrino oscillations probabilities. The procedure for deriving those equations is given here.

### A.1.1 The case of normal hierarchy of neutrino mass (NO)

In this case, the matrix $U_{PMNS}^{PDG}$ is used [3]

$$U_{PMNS}^{PDG} = \begin{pmatrix} U_{e1} & U_{e2} & U_{e3} \\ U_{\mu 1} & U_{\mu 2} & U_{\mu 3} \\ U_{\tau 1} & U_{\tau 2} & U_{\tau 3} \end{pmatrix} = \begin{pmatrix} c_{12}c_{13} & s_{12}c_{13} & s_{13}e^{-i\delta_{CP}} \\ -s_{12}c_{23}-c_{12}s_{23}s_{13}e^{i\delta_{CP}} & c_{12}c_{23}-s_{12}s_{23}s_{13}e^{i\delta_{CP}} & c_{13}s_{23} \\ s_{12}s_{23}-c_{12}c_{23}s_{13}e^{i\delta_{CP}} & -c_{12}s_{23}-s_{12}c_{23}s_{13}e^{i\delta_{CP}} & c_{13}c_{23} \end{pmatrix} = \begin{pmatrix} U_{e1} & U_{e2} & Je^{-i\delta_{CP}} \\ -A-Be^{i\delta_{CP}} & C-De^{i\delta_{CP}} & U_{\mu 3} \\ E-Fe^{i\delta_{CP}} & -G-He^{i\delta_{CP}} & U_{\tau 3} \end{pmatrix} \quad (A1)$$

where the mixing angles from the (A1) are taken into consideration $c_{ij} = \cos\theta_{ij}, s_{ij} = \sin\theta_{ij}; i,j = 1,2,3$.

In order to obtain an explicit numerical value of $\delta_{CP}$, the following unconditional rule will be applied: The sum of the probabilities of three neutrino oscillations during the transition $\nu_e \to \nu_\mu, \nu_e \to \nu_\tau, \nu_e \to \nu_e$, at a distance from the source equal to the entire wavelength of oscillations in the value of $X = L_{12}$, during the process of the disappearance in transition $\nu_e \to \nu_\mu \to \nu_e$, in the propagation of the neutrino beam through vacuum (as it can be seen, the matter effect is excluded in these considerations), is equal to one.

### A.1.2 Neutrino motion equation

In our considerations, we will use the general formula for neutrino oscillations given in [10] [12]:

$$P(\nu_\alpha \to \nu_\beta) = \delta_{\alpha\beta} - 4\sum_{i<j} Re\left(U_{\alpha i}U_{\beta i}^* U_{\alpha j}^* U_{\beta j}\right)\sin^2\left(\frac{\Delta m_{ji}^2 Lc^3}{4E\hbar}\right) + 2\sum_{i<j} Im\left(U_{\alpha i}U_{\beta i}^* U_{\alpha j}^* U_{\beta j}\right)\sin\left(\frac{\Delta m_{ji}^2 Lc^3}{2E\hbar}\right); i,j=1,2,3. \quad (A2)$$

**The transition is analysed:** $\nu_e \to \nu_\mu, \nu_e \to \nu_\tau, \nu_e \to \nu_e$

On the basis of formulae (A2), the total probability of neutrino oscillations is shown through the equation [17]

$$P(\nu_e \to \nu_\mu) + P(\nu_e \to \nu_\tau) + P(\nu_e \to \nu_e) = 1 - 4R\left\{U_{e1}U^*_{\mu 1}U^*_{e2}U_{\mu 2}\sin^2\pi\frac{\Delta m^2_{21}}{\Delta m^2_{21}}\right\} + 2\mathrm{Im}\left\{U_{e1}U^*_{\mu 1}U^*_{e2}U_{\mu 2}\sin 2\pi\frac{\Delta m^2_{21}}{\Delta m^2_{21}}\right\} - 4R\left\{U_{e1}U^*_{\mu 1}U^*_{e3}U_{\mu 3}\sin^2\pi\frac{\Delta m^2_{31}}{\Delta m^2_{21}}\right\} + 2\mathrm{Im}\left\{U_{e1}U^*_{\mu 1}U^*_{e3}U_{\mu 3}\sin 2\pi\frac{\Delta m^2_{31}}{\Delta m^2_{21}}\right\}$$
$$- 4R\left\{U_{e2}U^*_{\mu 2}U^*_{e3}U_{\mu 3}\sin^2\pi\frac{\Delta m^2_{32}}{\Delta m^2_{21}}\right\} + 2\mathrm{Im}\left\{U_{e2}U^*_{\mu 2}U^*_{e3}U_{\mu 3}\sin 2\pi\frac{\Delta m^2_{32}}{\Delta m^2_{21}}\right\} - 4R\left\{U_{e1}U^*_{\tau 1}U^*_{e2}U_{\tau 2}\sin^2\pi\frac{\Delta m^2_{21}}{\Delta m^2_{21}}\right\} + 2\mathrm{Im}\left\{U_{e1}U^*_{\tau 1}U^*_{e2}U_{\tau 2}\sin 2\pi\frac{\Delta m^2_{21}}{\Delta m^2_{21}}\right\} - 4R\left\{U_{e1}U^*_{\tau 1}U^*_{e3}U_{\tau 3}\sin^2\pi\frac{\Delta m^2_{31}}{\Delta m^2_{21}}\right\} + 2\mathrm{Im}\left\{U_{e1}U^*_{\tau 1}U^*_{e3}U_{\tau 3}\sin 2\pi\frac{\Delta m^2_{31}}{\Delta m^2_{21}}\right\} \quad \text{(A3)}$$
$$- 4R\left\{U_{e2}U^*_{\tau 2}U^*_{e3}U_{\tau 3}\sin^2\pi\frac{\Delta m^2_{32}}{\Delta m^2_{21}}\right\} + 2\mathrm{Im}\left\{U_{e2}U^*_{\tau 2}U^*_{e3}U_{\tau 3}\sin 2\pi\frac{\Delta m^2_{32}}{\Delta m^2_{21}}\right\} - 4|U_{e1}|^2|U_{e2}|^2\sin^2\pi\frac{\Delta m^2_{21}}{\Delta m^2_{21}} - 4|U_{e1}|^2|U_{e3}|^2\sin^2\pi\frac{\Delta m^2_{31}}{\Delta m^2_{21}} - 4|U_{e2}|^2|U_{e3}|^2\sin^2\pi\frac{\Delta m^2_{32}}{\Delta m^2_{21}} = 1$$

And, from the equation (A3), the equation of neutrino motion is formed with a condition that the travelled distance of the neutrino beam, moving through a vacuum from the source, equals the neutrino wavelength $X = L_{12}$. So, it can be written as

$$-4R\left\{U_{e1}U^*_{\mu 1}U^*_{e2}U_{\mu 2}\sin^2\pi\frac{\Delta m^2_{21}}{\Delta m^2_{21}}\right\} + 2\mathrm{Im}\left\{U_{e1}U^*_{\mu 1}U^*_{e2}U_{\mu 2}\sin 2\pi\frac{\Delta m^2_{21}}{\Delta m^2_{21}}\right\} - 4R\left\{U_{e1}U^*_{\mu 1}U^*_{e3}U_{\mu 3}\sin^2\pi\frac{\Delta m^2_{31}}{\Delta m^2_{21}}\right\} + 2\mathrm{Im}\left\{U_{e1}U^*_{\mu 1}U^*_{e3}U_{\mu 3}\sin 2\pi\frac{\Delta m^2_{31}}{\Delta m^2_{21}}\right\} - 4R\left\{U_{e2}U^*_{\mu 2}U^*_{e3}U_{\mu 3}\sin^2\pi\frac{\Delta m^2_{32}}{\Delta m^2_{21}}\right\} + 2\mathrm{Im}\left\{U_{e2}U^*_{\mu 2}U^*_{e3}U_{\mu 3}\sin 2\pi\frac{\Delta m^2_{32}}{\Delta m^2_{21}}\right\}$$
$$- 4R\left\{U_{e1}U^*_{\tau 1}U^*_{e2}U_{\tau 2}\sin^2\pi\frac{\Delta m^2_{21}}{\Delta m^2_{21}}\right\} + 2\mathrm{Im}\left\{U_{e1}U^*_{\tau 1}U^*_{e2}U_{\tau 2}\sin 2\pi\frac{\Delta m^2_{21}}{\Delta m^2_{21}}\right\} - 4R\left\{U_{e1}U^*_{\tau 1}U^*_{e3}U_{\tau 3}\sin^2\pi\frac{\Delta m^2_{31}}{\Delta m^2_{21}}\right\} + 2\mathrm{Im}\left\{U_{e1}U^*_{\tau 1}U^*_{e3}U_{\tau 3}\sin 2\pi\frac{\Delta m^2_{31}}{\Delta m^2_{21}}\right\} - 4R\left\{U_{e2}U^*_{\tau 2}U^*_{e3}U_{\tau 3}\sin^2\pi\frac{\Delta m^2_{32}}{\Delta m^2_{21}}\right\} + 2\mathrm{Im}\left\{U_{e2}U^*_{\tau 2}U^*_{e3}U_{\tau 3}\sin 2\pi\frac{\Delta m^2_{32}}{\Delta m^2_{21}}\right\} \quad \text{(A4)}$$
$$- 4|U_{e1}|^2|U_{e2}|^2\sin^2\pi\frac{\Delta m^2_{21}}{\Delta m^2_{21}} - 4|U_{e1}|^2|U_{e3}|^2\sin^2\pi\frac{\Delta m^2_{31}}{\Delta m^2_{21}} - 4|U_{e2}|^2|U_{e3}|^2\sin^2\pi\frac{\Delta m^2_{32}}{\Delta m^2_{21}} = 0$$

Taking into account

$$V = \sin\left(2\pi\frac{\Delta m^2_{31}}{\Delta m^2_{21}}\right), V = \sin\left(2\pi\frac{\Delta m^2_{32}}{\Delta m^2_{21}}\right), W = \sin^2\left(\pi\frac{\Delta m^2_{31}}{\Delta m^2_{21}}\right), W = \sin^2\left(\pi\frac{\Delta m^2_{32}}{\Delta m^2_{21}}\right). \quad \text{(A5)}$$

the complex structure of equations (A4) is reduced to the form:

$$4WJ[U_{e1}(AU_{\mu 3} - EU_{\tau 3}) - U_{e2}(CU_{\mu 3} - GU_{\tau 3})]\cos\delta_{CP} + 2VJ[U_{e1}(U_{\tau 3}E - U_{\mu 3}A) + U_{e2}(U_{\mu 3}C - U_{\tau 3}G)]\sin\delta_{CP} - 4WU^2_{e1}J^2 - 3U^2_{e2}J^2 + 4WJ(BU_{\mu 3} + FU_{\tau 3}) + 4WJ(DU_{\mu 3} + HU_{\tau 3})$$
$$= (4WJ\cos\delta_{CP} - 2VJ\sin\delta_{CP})(U_{e1}AU_{\mu 3} - U_{e1}U_{\tau 3}E + U_{e2}CU_{\mu 3} - U_{e2}GU_{\tau 3}) - 4WU^2_{e1}J^2 - 3U^2_{e2}J^2 + 4WJU_{e1}(BU_{\mu 3} + FU_{\tau 3}) + 4WJU_{e2}(DU_{\mu 3} + HU_{\tau 3}) = 0 \quad \text{(A6)}$$

And this structure is reduced to an extremely simple form:

$$(4WJ \cos\delta_{CP} - 2VJ \sin\delta_{CP})\varsigma - \xi = 0 \tag{A7}$$

In this equation, the following expressions equal zero:

$$\varsigma = (U_{e1}AU_{\mu3} - U_{e1}EU_{\tau3} - U_{e2}CU_{\mu3} + U_{e2}GU_{\tau3}) = 0 \tag{A8}$$

Because

$$U_{\mu3}A - U_{\tau3}E = S_{23}C_{13} \times S_{12}C_{23} - C_{23}C_{13} \times S_{12}S_{23} = 0 \tag{A9}$$

$$U_{\tau3}G - U_{\mu3}C = C_{23}C_{13} \times C_{12}S_{23} - S_{23}C_{13} \times C_{12}S_{23} = 0 \tag{A10}$$

And

$$\xi = -4WU_{e1}^2 J^2 - 4WU_{e2}^2 J^2 + 4WU_{e1}J(BU_{\mu3} + FU_{\tau3}) + 4WU_{e2}J(DU_{\mu3} + HU_{\tau3}) = 0 \tag{A11}$$

Because

$$U_{\mu3}B + U_{\tau3}F - U_{e1}J = S_{23}C_{13} \times C_{12}S_{23}S_{13} + C_{23}C_{13} \times C_{12}C_{23}S_{13} - C_{12}C_{13}S_{13} = 0 \tag{A12}$$

$$U_{\mu3}D + U_{\tau3}H - U_{e2}J = S_{23}C_{13} \times S_{12}S_{23}S_{13} + C_{23}C_{13} \times S_{12}C_{23}S_{13} - S_{12}C_{13}S_{13} = 0 \tag{A13}$$

### A.1.3 The case of inverted hierarchy of neutrino masses (IO)

And, in this case, the matrix $U_{PMNS}^{PDG}$ is used [2] [4] [10] [12]

$$U_{PMNS}^{PDG} = \begin{pmatrix} U_{e1} & U_{e2} & U_{e3} \\ U_{\mu1} & U_{\mu2} & U_{\mu3} \\ U_{\tau1} & U_{\tau2} & U_{\tau3} \end{pmatrix} = \begin{pmatrix} c_{12}c_{13} & s_{12}c_{13} & s_{13}e^{-i\delta_{CP}} \\ -s_{12}c_{23} - c_{12}s_{23}s_{13}e^{i\delta_{CP}} & c_{12}c_{23} - s_{12}s_{23}s_{13}e^{i\delta_{CP}} & c_{13}s_{23} \\ s_{12}s_{23} - c_{12}c_{23}s_{13}e^{i\delta_{CP}} & -c_{12}s_{23} - s_{12}c_{23}s_{13}e^{i\delta_{CP}} & c_{13}c_{23} \end{pmatrix} = \begin{pmatrix} U_{e1} & U_{e2} & Je^{-i\delta_{CP}} \\ -A - Be^{i\delta_{CP}} & C - De^{i\delta_{CP}} & U_{\mu3} \\ E - Fe^{i\delta_{CP}} & -G - He^{i\delta_{CP}} & U_{\tau3} \end{pmatrix} \tag{A14}$$

In order to obtain an explicit numerical value of $\delta_{CP}$, the following unconditional rule will be applied: The sum of the probabilities of three neutrino oscillations during the transition $\nu_e \to \nu_\mu, \nu_e \to \nu_\tau, \nu_e \to \nu_e$, at a distance from the source equal to the entire wavelength of oscillations in the value of $X = L_{12}$, during the process of the disappearance in transition $\nu_e \to \nu_\mu \to \nu_e$, in the propagation of the neutrino beam through vacuum (as it can be seen, the matter effect is excluded in these considerations), is equal to one.

$$P(\nu_e \to \nu_\mu) + P(\nu_e \to \nu_\tau) + P(\nu_e \to \nu_e) = 1 - 4R\left\{U_{e1}U^*_{\mu 1}U^*_{e2}U_{\mu 2}\sin^2\pi\frac{\Delta m^2_{21}}{\Delta m^2_{21}}\right\} + 2\operatorname{Im}\left\{U_{e1}U^*_{\mu 1}U^*_{e2}U_{\mu 2}\sin 2\pi\frac{\Delta m^2_{21}}{\Delta m^2_{21}}\right\} - 4R\left\{U_{e1}U^*_{\mu 1}U^*_{e3}U_{\mu 3}\sin^2\pi\frac{\Delta m^2_{13}}{\Delta m^2_{21}}\right\} + 2\operatorname{Im}\left\{U_{e1}U^*_{\mu 1}U^*_{e3}U_{\mu 3}\sin 2\pi\frac{\Delta m^2_{13}}{\Delta m^2_{21}}\right\}$$
$$- 4R\left\{U_{e2}U^*_{\mu 2}U^*_{e3}U_{\mu 3}\sin^2\pi\frac{\Delta m^2_{23}}{\Delta m^2_{21}}\right\} + 2\operatorname{Im}\left\{U_{e2}U^*_{\mu 2}U^*_{e3}U_{\mu 3}\sin 2\pi\frac{\Delta m^2_{23}}{\Delta m^2_{21}}\right\} - 4R\left\{U_{e1}U^*_{\tau 1}U^*_{e2}U_{\tau 2}\sin^2\pi\frac{\Delta m^2_{21}}{\Delta m^2_{21}}\right\} + 2\operatorname{Im}\left\{U_{e1}U^*_{\tau 1}U^*_{e2}U_{\tau 2}\sin 2\pi\frac{\Delta m^2_{21}}{\Delta m^2_{21}}\right\} - 4R\left\{U_{e1}U^*_{\tau 1}U^*_{e3}U_{\tau 3}\sin^2\pi\frac{\Delta m^2_{13}}{\Delta m^2_{21}}\right\}$$
$$+ 2\operatorname{Im}\left\{U_{e1}U^*_{\tau 1}U^*_{e3}U_{\tau 3}\sin 2\pi\frac{\Delta m^2_{13}}{\Delta m^2_{21}}\right\} - 4R\left\{U_{e2}U^*_{\tau 2}U^*_{e3}U_{\tau 3}\sin^2\pi\frac{\Delta m^2_{23}}{\Delta m^2_{21}}\right\} + 2\operatorname{Im}\left\{U_{e2}U^*_{\tau 2}U^*_{e3}U_{\tau 3}\sin 2\pi\frac{\Delta m^2_{23}}{\Delta m^2_{21}}\right\} - 4|U_{e1}|^2|U_{e2}|^2\sin^2\pi\frac{\Delta m^2_{21}}{\Delta m^2_{21}} - 4|U_{e1}|^2|U_{e3}|^2\sin^2\pi\frac{\Delta m^2_{13}}{\Delta m^2_{21}} - 4|U_{e2}|^2|U_{e3}|^2\sin^2\pi\frac{\Delta m^2_{23}}{\Delta m^2_{21}} = 1$$

(A15)

From the equation (A15), the following form of the motion equation ensues [17]:

$$-4R\left\{U_{e1}U^*_{\mu 1}U^*_{e2}U_{\mu 2}\sin^2\pi\frac{\Delta m^2_{21}}{\Delta m^2_{21}}\right\} + 2\operatorname{Im}\left\{U_{e1}U^*_{\mu 1}U^*_{e2}U_{\mu 2}\sin 2\pi\frac{\Delta m^2_{21}}{\Delta m^2_{21}}\right\} - 4R\left\{U_{e1}U^*_{\mu 1}U^*_{e3}U_{\mu 3}\sin^2\pi\frac{\Delta m^2_{13}}{\Delta m^2_{21}}\right\} + 2\operatorname{Im}\left\{U_{e1}U^*_{\mu 1}U^*_{e3}U_{\mu 3}\sin 2\pi\frac{\Delta m^2_{13}}{\Delta m^2_{21}}\right\} - 4R\left\{U_{e2}U^*_{\mu 2}U^*_{e3}U_{\mu 3}\sin^2\pi\frac{\Delta m^2_{23}}{\Delta m^2_{21}}\right\}$$
$$+ 2\operatorname{Im}\left\{U_{e2}U^*_{\mu 2}U^*_{e3}U_{\mu 3}\sin 2\pi\frac{\Delta m^2_{23}}{\Delta m^2_{21}}\right\} - 4R\left\{U_{e1}U^*_{\tau 1}U^*_{e2}U_{\tau 2}\sin^2\pi\frac{\Delta m^2_{21}}{\Delta m^2_{21}}\right\} + 2\operatorname{Im}\left\{U_{e1}U^*_{\tau 1}U^*_{e2}U_{\tau 2}\sin 2\pi\frac{\Delta m^2_{21}}{\Delta m^2_{21}}\right\} - 4R\left\{U_{e1}U^*_{\tau 1}U^*_{e3}U_{\tau 3}\sin^2\pi\frac{\Delta m^2_{13}}{\Delta m^2_{21}}\right\} + 2\operatorname{Im}\left\{U_{e1}U^*_{\tau 1}U^*_{e3}U_{\tau 3}\sin 2\pi\frac{\Delta m^2_{13}}{\Delta m^2_{21}}\right\}$$
$$- 4R\left\{U_{e2}U^*_{\tau 2}U^*_{e3}U_{\tau 3}\sin^2\pi\frac{\Delta m^2_{23}}{\Delta m^2_{21}}\right\} + 2\operatorname{Im}\left\{U_{e2}U^*_{\tau 2}U^*_{e3}U_{\tau 3}\sin 2\pi\frac{\Delta m^2_{23}}{\Delta m^2_{21}}\right\} - 4|U_{e1}|^2|U_{e2}|^2\sin^2\pi\frac{\Delta m^2_{21}}{\Delta m^2_{21}} - 4|U_{e1}|^2|U_{e3}|^2\sin^2\pi\frac{\Delta m^2_{13}}{\Delta m^2_{21}} - 4|U_{e2}|^2|U_{e3}|^2\sin^2\pi\frac{\Delta m^2_{23}}{\Delta m^2_{21}} = 0$$

(A16)

Taking into account

$$V = \sin\left(2\pi\frac{\Delta m^2_{13}}{\Delta m^2_{21}}\right), V = \sin\left(2\pi\frac{\Delta m^2_{23}}{\Delta m^2_{21}}\right), W = \sin^2\left(\pi\frac{\Delta m^2_{13}}{\Delta m^2_{21}}\right), W = \sin^2\left(\pi\frac{\Delta m^2_{23}}{\Delta m^2_{21}}\right)$$

(A17)

the equation (A16) is reduced to the form:

$$4WJ[U_{e1}(AU_{\mu3} - EU_{\tau3}) - U_{e2}(CU_{\mu3} - GU_{\tau3})]\cos\delta_{CP} + 2VJ[U_{e1}(U_{\tau3}E - U_{\mu3}A) + U_{e2}(U_{\mu3}C - U_{\tau3}G)]\sin\delta_{CP} - 4WU_{e1}^2 J^2 - 4WU_{e2}^2 J^2 + 4WJ(BU_{\mu3} + FU_{\tau3}) + 4WJ(DU_{\mu3} + HU_{\tau3})$$
$$= (4WJ\cos\delta_{CP} - 2VJ\sin\delta_{CP})(U_{e1}AU_{\mu3} - U_{e1}U_{\tau3}E + U_{e2}CU_{\mu3} - U_{e2}GU_{\tau3}) - 4WU_{e1}^2 J^2 - 3U_{e2}^2 J^2 + 4WJU_{e1}(BU_{\mu3} + FU_{\tau3}) + 4WJU_{e2}(DU_{\mu3} + HU_{\tau3}) = 0 \quad (A18)$$

And this complex structure is reduced to an extremely simple form:

$$(4WJ\cos\delta_{CP} - 2VJ\sin\delta_{CP})\varsigma - \xi = 0 \quad (A19)$$

In this equation, the following expressions equal zero:

$$\varsigma = (U_{e1}AU_{\mu3} - U_{e1}EU_{\tau3} - U_{e2}CU_{\mu3} + U_{e2}GU_{\tau3}) = 0 \quad (A20)$$

Because

$$U_{\mu3}A - U_{\tau3}E = S_{23}C_{13} \times S_{12}C_{23} - C_{23}C_{13} \times S_{12}S_{23} = 0 \quad (A21)$$

$$U_{\tau3}G - U_{\mu3}C = C_{23}C_{13} \times C_{12}S_{23} - S_{23}C_{13} \times C_{12}S_{23} = 0 \quad (A22)$$

And

$$\xi = -4WU_{e1}^2 J^2 - 4WU_{e2}^2 J^2 + 4WU_{e1}J(BU_{\mu3} + FU_{\tau3}) + 4WU_{e2}J(DU_{\mu3} + HU_{\tau3}) = 0 \quad (A23)$$

Because

$$U_{\mu3}B + U_{\tau3}F - U_{e1}J = S_{23}C_{13} \times C_{12}S_{23}S_{13} + C_{23}C_{13} \times C_{12}C_{23}S_{13} - C_{12}C_{13}S_{13} = 0 \quad (A24)$$

$$U_{\mu3}D + U_{\tau3}H - U_{e2}J = S_{23}C_{13} \times S_{12}S_{23}S_{13} + C_{23}C_{13} \times S_{12}C_{23}S_{13} - S_{12}C_{13}S_{13} = 0 \quad (A25)$$

Additionally to that and based on the results (A8, A11, A20, A23), it could be concluded that the Dirac CP violation phase could not depend on the mixing angles: $\theta_{12}, \theta_{13}, \theta_{23}$, and the equation (A19) is reduced to the form:

$$(4W\cos\delta_{CP} - 2V\sin\delta_{CP})*0 = 0 \tag{A26}$$

The first point that can be stated is that this equation is always satisfied for any value of $\delta_{CP} \in [0,2\pi)$, so such solutions make no physical sense. It is apparent that among those solutions in the range $\delta_{CP} \in [0,2\pi)$ there is the right unique solution for the value $\delta_{CP}$. From such set of countless values, the real and unique value for $\delta_{CP}$ is drawn from the set $\delta_{CP} \in [0,2\pi)$ by solving the equation (A26)

$$2W\cos\delta_{CP} - V\sin\delta_{CP} = 0 \tag{A27}$$

The solution of this equation presents the particular solution of the equation (A26), and it is in the following form:

*Normal ordering*

$$\delta CP(NO) = arctg\left(\frac{2W}{V}\right) = arctg\left[2\frac{\sin^2\left(\pi\frac{\Delta m_{31}^2}{\Delta m_{21}^2}\right)}{\sin\left(2\pi\frac{\Delta m_{31}^2}{\Delta m_{21}^2}\right)}\right] = arctg\left[tg\left(\pi\frac{\Delta m_{31}^2}{\Delta m_{21}^2}\right)\right] = \langle\delta CP(NO)\rangle + |\Delta(\delta CP(NO))| = 180^0 \times \frac{\langle\Delta m_{31}^2\rangle}{\langle\Delta m_{21}^2\rangle} + \left|\Delta\left(180^0 \times \frac{\Delta m_{31}^2}{\Delta m_{21}^2}\right)\right|,$$

$$\Delta m_{31}^2 = \langle\Delta m_{31}^2\rangle \pm \Delta(\Delta m_{31}^2), \Delta m_{21}^2 = \langle\Delta m_{21}^2\rangle \pm \Delta(\Delta m_{21}^2)$$

(A28)

*Inverted ordering*

$$\delta CP(IO) = arctg\left(\frac{2W}{V}\right) = arctg\left[2\frac{\sin^2\left(\pi\frac{\Delta m_{23}^2}{\Delta m_{21}^2}\right)}{\sin\left(2\pi\frac{\Delta m_{23}^2}{\Delta m_{21}^2}\right)}\right] = arctg\left[tg\left(\pi\frac{\Delta m_{23}^2}{\Delta m_{21}^2}\right)\right] = \langle\delta CP(IO)\rangle + |\Delta(\delta CP(IO))| = 180^0 \times \frac{\langle\Delta m_{23}^2\rangle}{\langle\Delta m_{21}^2\rangle} + \left|\Delta\left(180^0 \times \frac{\Delta m_{23}^2}{\Delta m_{21}^2}\right)\right|,$$

$$\Delta m_{23}^2 = \langle\Delta m_{23}^2\rangle \pm \Delta(\Delta m_{23}^2), \Delta m_{21}^2 = \langle\Delta m_{21}^2\rangle \pm \Delta(\Delta m_{21}^2)$$

(A29)